\documentclass[twocolumn]{aastex62}

\usepackage{mathptmx}
\usepackage[rightcaption]{sidecap}

\usepackage{nicefrac}

\definecolor{myGreen}{cmyk}{0.992,0.,0.083, 0.525}

\definecolor{myDarkRed}{rgb}{0.698, 0.094, 0.133}

\received{\today}
\revised{later}
\accepted{later}

\submitjournal{The Astrophysical Journal}

\shorttitle{Classification of High-resolution Solar H$\alpha$ Spectra}
\shortauthors{Verma et al.}

\begin{document}


\title{Classification of High-resolution Solar H$\alpha$ Spectra using
    t-distributed Stochastic Neighbor Embedding}

\correspondingauthor{Meetu Verma}
\email{mverma@aip.de}
\author[0000-0003-1054-766X]{Meetu Verma}
\affil{Leibniz-Institut f\"ur Astrophysik Potsdam (AIP),
    An der Sternwarte 16, 14482 Potsdam, Germany}

\author[0000-0002-6070-2288]{Gal Matijevi{\v{c}}}
\affil{Leibniz-Institut f\"ur Astrophysik Potsdam (AIP),
    An der Sternwarte 16, 14482 Potsdam, Germany}

\author[0000-0002-7729-6415]{Carsten Denker}
\affil{Leibniz-Institut f\"ur Astrophysik Potsdam (AIP),
    An der Sternwarte 16, 14482 Potsdam, Germany}

\author[0000-0002-9858-0490]{Andrea Diercke}
\affil{Leibniz-Institut f\"ur Astrophysik Potsdam (AIP),
    An der Sternwarte 16, 14482 Potsdam, Germany}
\affiliation{Universit{\"a}t Potsdam, Institut f{\"u}r Physik und Astronomie,
    Karl-Liebknecht-Stra{\ss}e 24\,--\,25, 14476 Potsdam, Germany}

\author[0000-0002-4645-4492]{Ekaterina Dineva}
\affil{Leibniz-Institut f\"ur Astrophysik Potsdam (AIP),
    An der Sternwarte 16, 14482 Potsdam, Germany}
\affiliation{Universit{\"a}t Potsdam, Institut f{\"u}r Physik und Astronomie,
    Karl-Liebknecht-Stra{\ss}e 24\,--\,25, 14476 Potsdam, Germany}

\author[0000-0002-4739-1710]{Horst Balthasar}
\affil{Leibniz-Institut f\"ur Astrophysik Potsdam (AIP),
    An der Sternwarte 16, 14482 Potsdam, Germany}

\author[0000-0003-2059-585X]{Robert Kamlah}
\affil{Leibniz-Institut f\"ur Astrophysik Potsdam (AIP),
    An der Sternwarte 16, 14482 Potsdam, Germany}
\affiliation{Universit{\"a}t Potsdam, Institut f{\"u}r Physik und Astronomie,
    Karl-Liebknecht-Stra{\ss}e 24\,--\,25, 14476 Potsdam, Germany}

\author[0000-0002-3694-4527]{Ioannis Kontogiannis}
\affil{Leibniz-Institut f\"ur Astrophysik Potsdam (AIP),
    An der Sternwarte 16, 14482 Potsdam, Germany}

\author[0000-0002-3242-1497]{Christoph Kuckein}
\affil{Leibniz-Institut f\"ur Astrophysik Potsdam (AIP),
    An der Sternwarte 16, 14482 Potsdam, Germany}

\author[0000-0001-8670-2123]{Partha S. Pal}
\affil{Leibniz-Institut f\"ur Astrophysik Potsdam (AIP),
    An der Sternwarte 16, 14482 Potsdam, Germany}
\affiliation{University of Delhi, Bhaskaracharya College of Applied Sciences,
    Delhi 110007, India}


\begin{abstract}
The H$\alpha$ spectral line is a well-studied absorption line revealing properties of the highly structured and dynamic solar chromosphere. Typical features with distinct spectral signatures in H$\alpha$ include filaments and prominences, bright active-region plages, superpenumbrae around sunspots, surges, flares, Ellerman bombs, filigree, and mottles and rosettes, among others. This study is based on high-spectral resolution H$\alpha$ spectra obtained with the echelle spectrograph of the Vacuum Tower Telescope (VTT) located at Observatorio del Teide (ODT), Tenerife, Spain. The t-distributed Stochastic Neighbor Embedding (t-SNE) is a machine learning algorithm, which is used for nonlinear dimensionality reduction. In this application, it projects H$\alpha$ spectra onto a two-dimensional map, where it becomes possible to classify the spectra according to results of Cloud Model (CM) inversions. The CM parameters optical depth, Doppler width, line-of-sight velocity, and source function describe properties of the cloud material. Initial results of t-SNE indicate its strong discriminatory power to separate quiet-Sun and plage profiles from those that are suitable for CM inversions.  In addition, a detailed study of various t-SNE parameters
is conducted, the impact of seeing conditions on the classification is assessed, results for various types of input data are compared, and the identified clusters are linked to chromospheric features. Although t-SNE proves to be efficient in clustering high-dimensional data, human inference is required at each step to interpret the results. This exploratory study provides a framework and ideas on how to tailor a classification scheme towards specific spectral data and science questions. 

\end{abstract}



\section{Introduction} \label{sec:intro}

The volume and complexity of data is increasing in solar physics with the advent of new observing facilities and instrumentation. A typical three-dimensional data cube contains one wavelength and two spatial dimensions. However, photon-efficient instruments and fast cameras facilitate recording time-series with high cadence, adding time as a fourth dimension. Furthermore, whenever magnetic fields are observed, the polarization state of the light constitutes another dimension. Various linear and nonlinear techniques are proposed for dimensionality reduction with the aim to preserve the structure of the data. Stochastic Neighbor Embedding \citep[SNE,][]{Hinton2002} is one of these techniques, which performs well on artificial data while struggling in visualizing real-world, high-dimensional data. Therefore, \citet{vanderMaaten2008} proposed t-distributed Stochastic Neighbor Embedding (t-SNE), which transforms a high-dimensional dataset into a matrix of pairwise similarities. Thus, t-SNE captures the local structure and at the same time reveals the global structure of the dataset. The mathematical details of the technique (see Sect.~\ref{sec:methods}) were elaborated in \citet{vanderMaaten2008}, and implementations of the code in various programming languages are publicly available.\footnote{\href{https://lvdmaaten.github.io/tsne/}{lvdmaaten.github.io/tsne}}

t-SNE is a very powerful tool for the initial selection of data, thus reducing their dimensionality. \citet{Matijevic2017} employed t-SNE in three stages to detect metal poor stars in the complete survey of the Radial Velocity Experiment \citep[RAVE,][]{Steinmetz2006}. They used t-SNE to create a low-dimensional projection of the spectrum space and to segregate the metal-poor star region. The final projection of spectra enables them to detect classes of the stars with similar atmospheric parameters, in particular clusters of metal-poor stars, which were used in the subsequent in-depth spectral analysis.

\citet{Traven2017} used t-SNE to recognize different spectral features and to classify stellar spectra in the Galactic Archaeology with Hermes \citep[GALAH,][]{DeSilva2015} survey. About 77\,500 spectra were selected out of 210\,000 GALAH spectra using iteratively t-SNE and an algorithm for density-based spatial clustering \citep[DBSCAN,][]{Ester1996}. The classification procedure arrived at six classes of stellar spectra. Further work on the GALAH survey was carried out by \citet{Kos2018} who used t-SNE as a tool for chemical tagging and as a way to visualize data in high-dimensional space. From about 187\,000 stars with 13 chemical abundances, the t-SNE algorithm retrieved nine clusters in the chemical abundance space. The robustness of the t-SNE algorithm was further demonstrated by \citet{Anders2018} who visualized chemical abundances of stars contained in the HARPS-GTO exoplanet search program \citep{DelgadoMena2017}. The t-SNE algorithm allowed them to define more reliably the chemical sub-populations of solar-neighborhood stars. In addition, t-SNE projected maps revealed many stars with peculiar chemical compositions.

Solar physics as well as other branches of astronomy entered an era of rapidly growing data volumes, e.g., high-resolution imaging and spectroscopy, synoptic observations, and sophisticated numerical modeling of solar phenomena. Hence, machine learning techniques became mainstream and are used by the community to tackle ever more complex tasks. When analyzing spectral observations with machine learning techniques, spectral profiles serve as portals to information about the physical state of plasma at various levels of the solar atmosphere.

\citet{Carroll2001} examined three different strategies for the inversion of spectral lines and Stokes profiles using artificial neural networks (ANNs). They used the photospheric infrared Fe\,\textsc{i} line and estimated magnetic field and other physical parameters with remarkable speed. \citet{SocasNavarro2005c} carried out similar work with a three-stage approach for spectral line inversion based on ANNs. First, the network was trained with synthetic spectra from a simplified model. Second, additional preprocessing using autoassociative neural networks projected the observation to the theoretical model space. The third stage included regularization of the neural network. \citet{Carroll2008} expanded their previous work using multi-layer perceptrons (MLPs) to coordinate the training of their network with a quiet-Sun simulation to obtain three-dimensional information of temperature, velocity, and magnetic field vector. Recently, \citet{AsensioRamos2019} presented a fast inversion code using convolutional neural networks (CNNs). They trained two different CNN architectures, where they used synthetic Stokes profiles from two snapshots of a three-dimensional magnetohydrodynamic numerical simulations containing different structures of the solar atmosphere. For lines originating in solar transition region as observed with Interface Region Imaging Spectrograph \citep[IRIS,][]{DePontieu2014}, \citet{Dalda2019} presented a faster way to obtain thermodynamic properties by combining the results provided by the traditional methods with machine and deep learning techniques.

Although, machine learning is commonly used in solar spectral line inversions, these techniques are not yet extensively used for classification and identification of the spectral profiles. One of the few works is by \citet{Panos2018}, who employed supervised hierarchical k-means to identify typical Mg\,\textsc{ii} flare spectra. \citet{Panos2020} extended their previous work to real time prediction of solar flare by applying a Deep Neural Network (DNN) to Mg\,\textsc{ii} spectra. In addition to using neural networks (NNs), they employed principal component analysis (PCA) and t-SNE for low dimensional representation to examine the behaviour of various features. Furthermore, \citet{Kuckein2020} employed the unsupervised machine-learning algorithm k-means to classify He\,\textsc{i}~10830~\AA\, spectra, with the aim of significantly improving the quality and speed of spectral-line inversions of this triplet.

Various machine learning techniques have been used in solar physics but employing t-SNE to classify or identify clusters in spectral data is still very new \citep{Verma2019}. In this work, we present an exploratory study of t-SNE-based classification of high-resolution H$\alpha$ spectra. Data and methods are introduced in Sects.~\ref{sec:data} and~\ref{sec:methods}, respectively. Results, including a parameter study to find an optimal t-SNE setup, are described in Sect.~\ref{sec:results}. Discussions and an outlook are presented in Sects.~\ref{sec:discussion} and~\ref{sec:conclusions}, respectively.


\section{Observations and data} \label{sec:data}

\begin{figure}[t]
\centering
\includegraphics[width=\columnwidth]{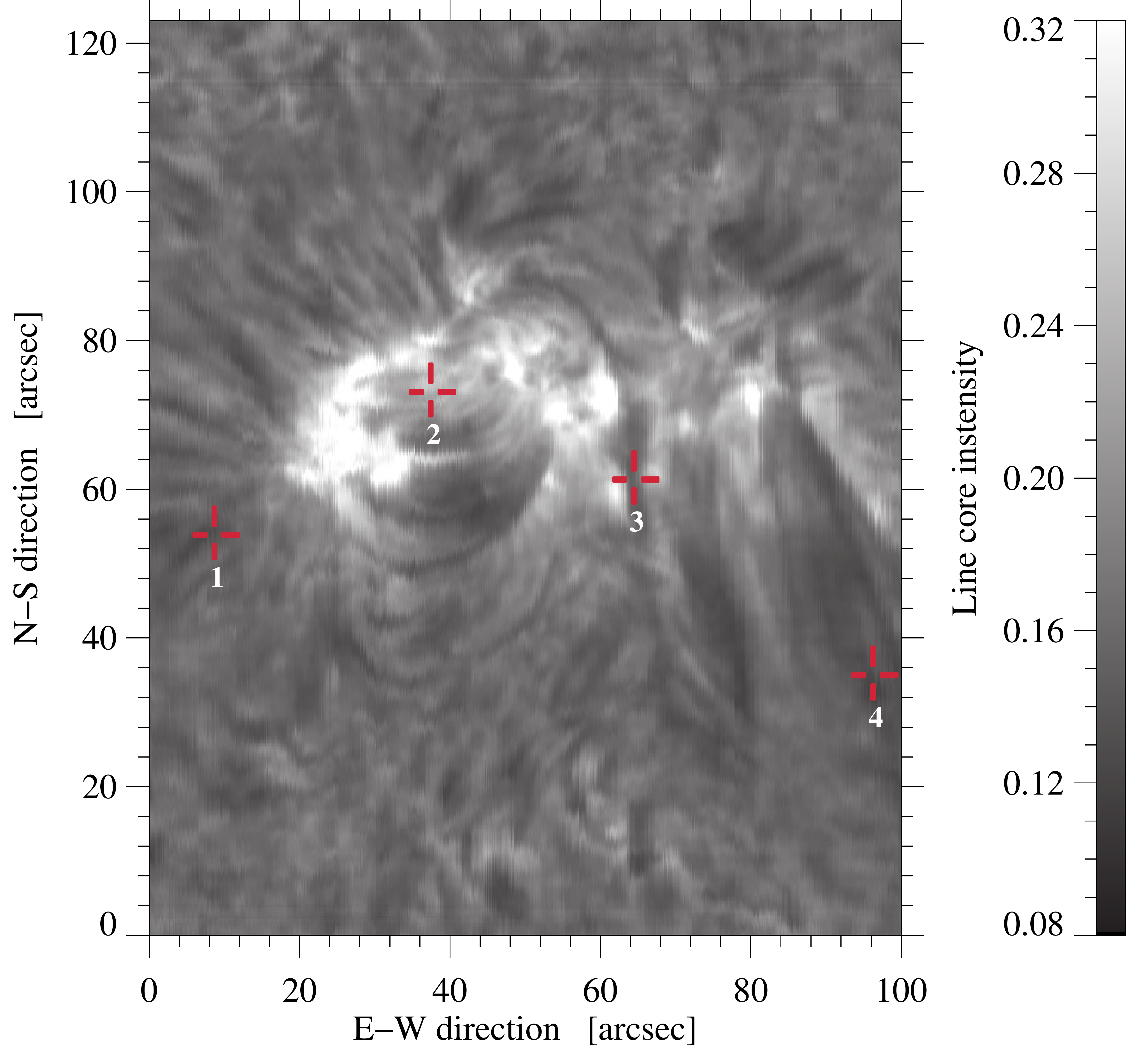}
\caption{Slit-reconstructed H$\alpha$ line-core intensity image of active region 
NOAA~12722 observed at 08:05 UT on 11 September 2018.}
\label{FIG02a}
\end{figure}

High-spectral resolution H$\alpha$ (6562.8 \AA) spectra were obtained on 11~September 2018 with the echelle spectrograph of the Vacuum Tower Telescope \citep[VTT,][]{vonderLuehe1998} at Observatorio del Teide (ODT), Tenerife, Spain. 
The details of observations and data processing are presented by \citet{Verma2020b}, who investigated a surge in active region NOAA~12722. This is the reference publication for the temporal evolution of the active region -- its morphology in photosphere, chromosphere, and upper atmosphere, its complex velocity fields associated with ejected and returning plasma, and its spectral characteristics of surging plasma. Starting at 08:05~UT, 21 H$\alpha$ spatio-spectral scans were acquired in three and half hours. Each scan took about 9~min with a few longer time gaps between some scans. The field-of-view (FOV) of $100\arcsec \times 120\arcsec$ is covered in each scan with 630 scan steps and 660~pixels along the slit as shown in the slit-reconstructed line-core intensity map in Fig.~\ref{FIG02a}. In the center of the FOV resides an arch-filament system connecting two opposite polarities. 
The size of a scan step is 0.16\arcsec\ and a pixel along the slit corresponds to 0.18\arcsec. The data processing steps were described in \citet{Dineva2020}, which included PCA for noise stripping and computation of Cloud Model (CM) inversions \citep{Beckers1964}. The final spectra are resampled to 601 wavelength points, which cover a wavelength range of $\pm$3~\AA\ around the H$\alpha$ line core. These high-spectral and moderate-spatial resolution spectra are the basis of this work.

\begin{figure}[t]
\centering
\includegraphics[width=\columnwidth]{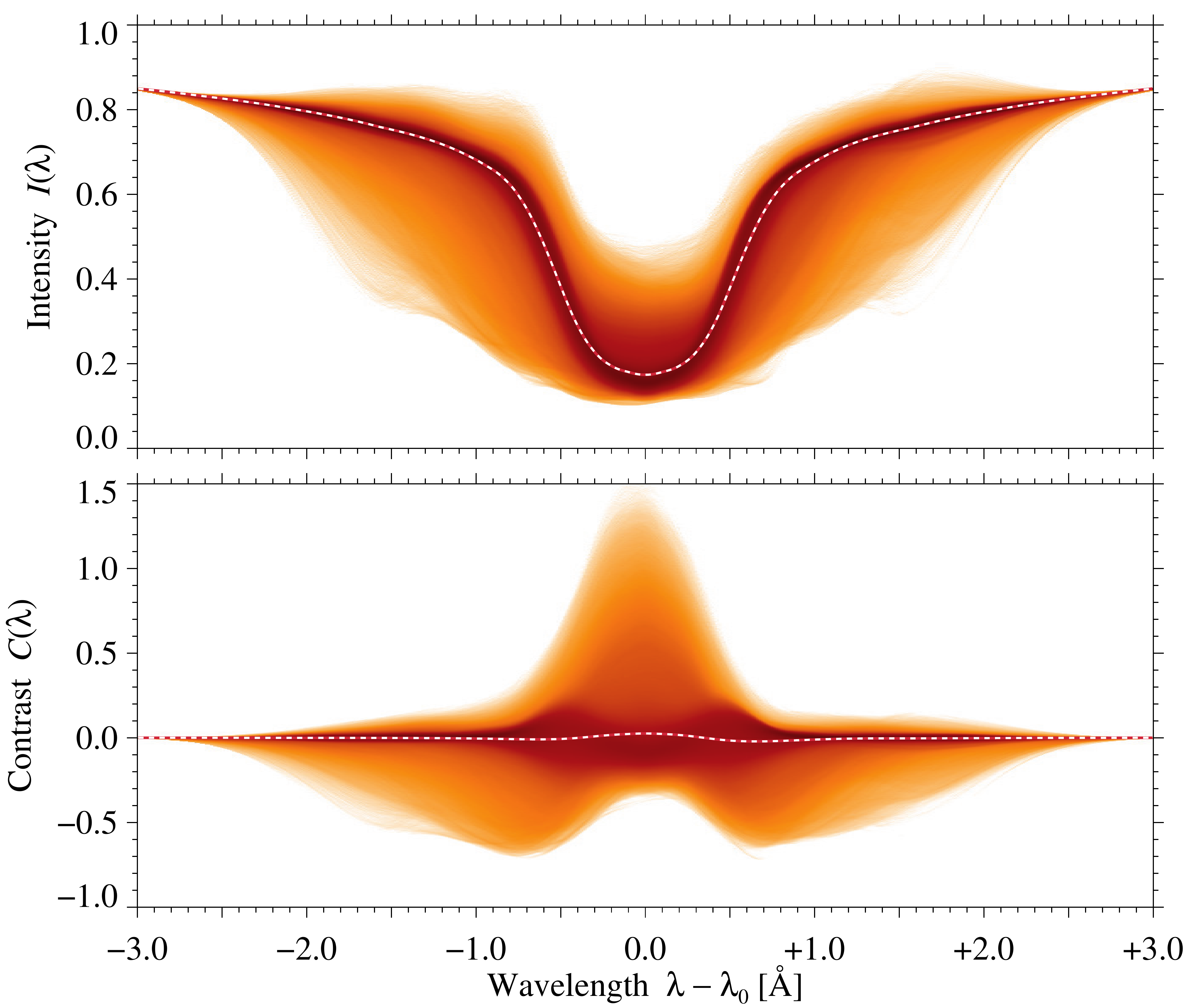}
\caption{Two-dimensional histograms of observed, noise-stripped H$\alpha$
    intensity (\textit{top}) and contrast (\textit{bottom}) profiles. The distributions were divided by the number of profiles (about 8.7 million) and are displayed on a logarithmic scale between $10^{-6}$ and $10^0$, where darker colors refer to a higher number density. The dashed curves in the top and bottom panels refer to the average H$\alpha$ intensity and contrast profiles, respectively.}
\label{FIG01}
\end{figure}

\begin{figure}[t]
\centering
\includegraphics[width=\columnwidth]{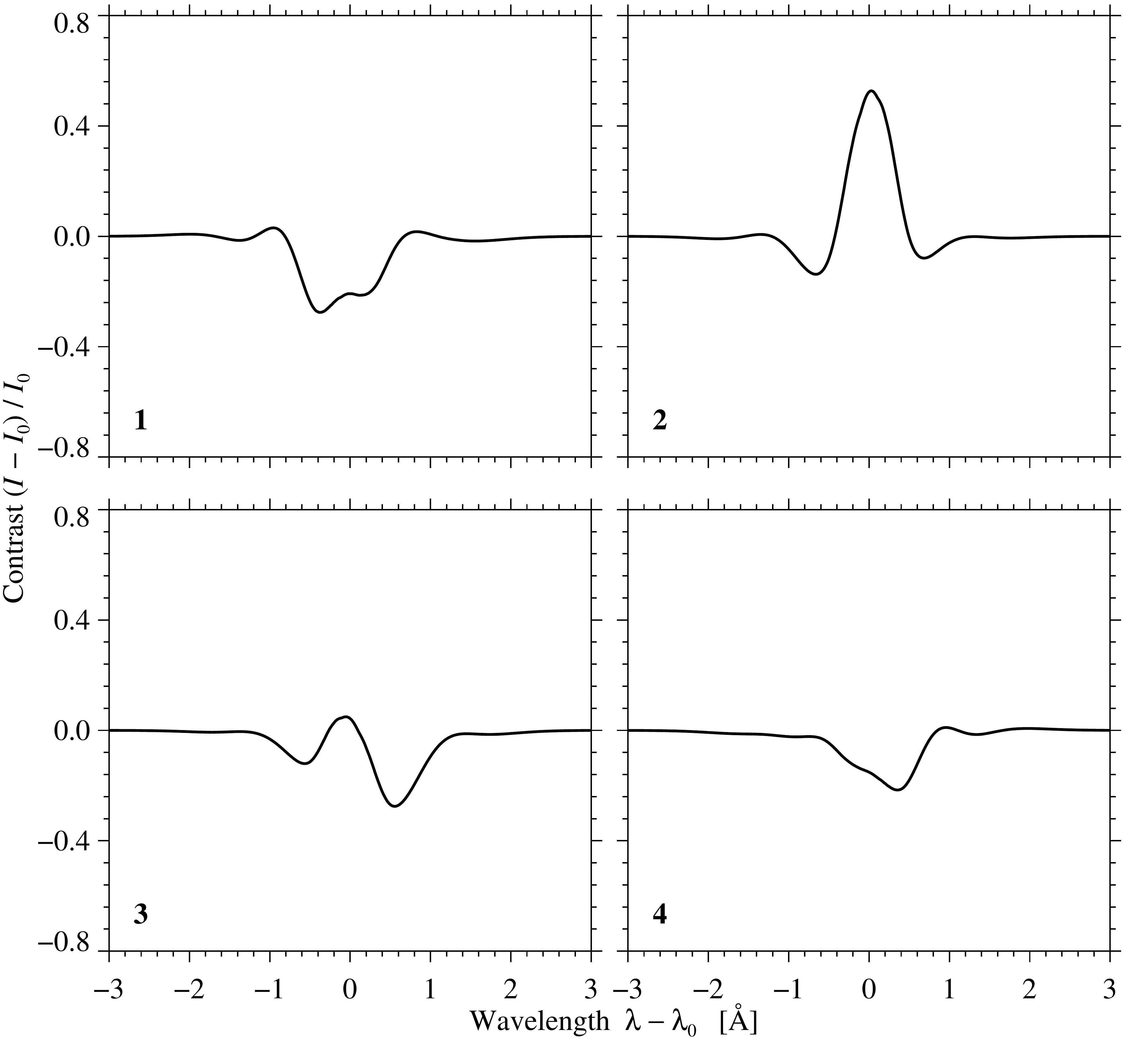}
\caption{Four noise-stripped contrast profiles, belonging to the locations
marked in Fig.~\ref{FIG02a}, which demonstrate the diversity of the observed profiles.}
\label{FIG01a}
\end{figure}

The time-series of 21 H$\alpha$ spatio-spectral data cubes contains about $2 \times 8.7$~million intensity and contrast profiles. The contrast profiles refer to 
\begin{equation}
C(\lambda) = \frac{I(\lambda) - I_{0}(\lambda)}{I_{0}(\lambda)},
\end{equation}
where $I_{0}(\lambda)$ is the quiet-Sun spectral profile obtained either from observations or model atmospheres. We used the observed quiet-Sun background profiles of \citet{David1961} and interpolated them for the heliocentric angle of $\mu = 0.86$ as explained in \citet{Verma2020b} and \citet{Dineva2020}. The morphology of the spectral shapes and their frequency of occurrence are visualized summarily in Fig.~\ref{FIG01}. The two-dimensional histograms have a bin size of 10~m\AA\ along the wavelength axis and $2 \times 10^{-3}$ and $5 \times 10^{-3}$ along the intensity and contrast axes, respectively. The histograms were normalized by the number of spectra and are displayed on a logarithmic scale, where light orange refers to about 10 spectral points per bin and the darkest orange indicates that all spectral points are contained in the bin. The Doppler shift of individual spectral line profiles was not corrected, which results in the extended wings of the two-dimensional histograms. In addition, the logarithmic display may be misleading as the vast majority of intensity profiles stays close to the average profile, which is depicted as a red-white dashed curve. The average contrast profile stays close to zero, while the frequency distribution indicates a preponderance of \textsf{W}-shaped profiles.

To demonstrate the variation in contrast profiles, we plotted four examples in Fig.~\ref{FIG01a}. The location of these four profiles are marked on the slit-reconstructed line-core intensity map in Fig.~\ref{FIG02a}. These profiles provide a glimpse of various contrast profiles ranging from contrast profiles  with a strong or weak central component (2 or 3) to contrast profiles, where the central maximum is less pronounced and the contrast is almost everywhere negative (1 and 4). The red-blue asymmetry is indicative of Doppler shifts.

CM inversions \citep{Beckers1964} are widely used in solar physics to
infer physical parameters for dark cloud-like absorption features observed in the H$\alpha$ line \citep[e.g.,][]{Tziotziou2007}. With some fundamental assumptions \citep[see e.g.,][]{Kuckein2016, Dineva2020} four parameters are determined which define the radiative transfer and line formation. These four parameters are the optical thickness $\tau_0$, Doppler velocity of the cloud $v_\mathrm{D}$, Doppler width $\Delta\lambda_\mathrm{D}$ of the absorption profile, and source function $S$. Profiles which are suitable for CM inversions are characterized by strong absorption profiles. Note that CM inversions cannot reproduce the strong central component with positive contrasts. More details on the implementation and computation of CM inversions are given in \citet{Dineva2020}.

In general, the seeing conditions were good at the beginning and deteriorated towards the end of the observing run. The best seeing conditions were encountered for scans No.~1, 2, 3, and 6, whereas the seeing conditions were worst for scans No.~20, 21, 9, and 15. Scan No.~1 with the best seeing conditions is used in the following for benchmarking the t-SNE algorithm. The seeing and image quality were determined from the granular contrast and the \textit{Median Filter Gradient Similarity} \citep[MFGS,][]{Deng2015, Denker2018b} of slit-reconstructed pseudo-continuum images. The granular contrast covered the range 1.56\,--\,2.14\%. The MFGS values reside by definition in the interval [0,\ 1], whereby higher values indicate better image quality and in turn better seeing conditions. Since the solar surface is scanned by the spectrograph, the seeing varies at each scanned step. However, visual inspection as well as MFGS values show that seeing estimates based on regions with granulation are representative for the full FOV. At the lowest contrast and MFGS values, the granular pattern is completely washed out and only the dark sunspot and pores as well as bright H$\alpha$ grains remain visible.

The benchmark spectral data comprise intensity profiles, contrast profiles, and their counterparts based on the superposition of specific eigenfunctions derived from PCA and CM inversions. According to \citet{Dineva2020}, ten eigenfunctions are sufficient to construct the observed H$\alpha$ profiles. In the present study, application of PCA and CM inversions yields noise-stripped contrast profiles, which are used as input for the t-SNE algorithm unless stated otherwise.


\section{Methods} \label{sec:methods}

Many machine learning algorithms are available for dimension reduction and visualization of multidimensional data. In our application, i.e., finding clusters or classes of high-resolution chromospheric spectra, t-SNE was chosen because it delivered very good results in classifying stellar spectra \citep{Matijevic2017}. The method was proposed by \citet{vanderMaaten2008} as a successor of SNE \citep{Hinton2002}. In this study, we primarily used contrast profiles of scans with good seeing, which contain $n_s = 660 \times 630 = 415\,800$ samples and $n_f = 601$ features (wavelength points). In terms of a multidimensional space, we have 415\,800 points in a 601-dimensional space. Our task is to evaluate the level of similarity between these points. 

\begin{figure*}[t]
\centering
\includegraphics[width=0.28\textwidth]{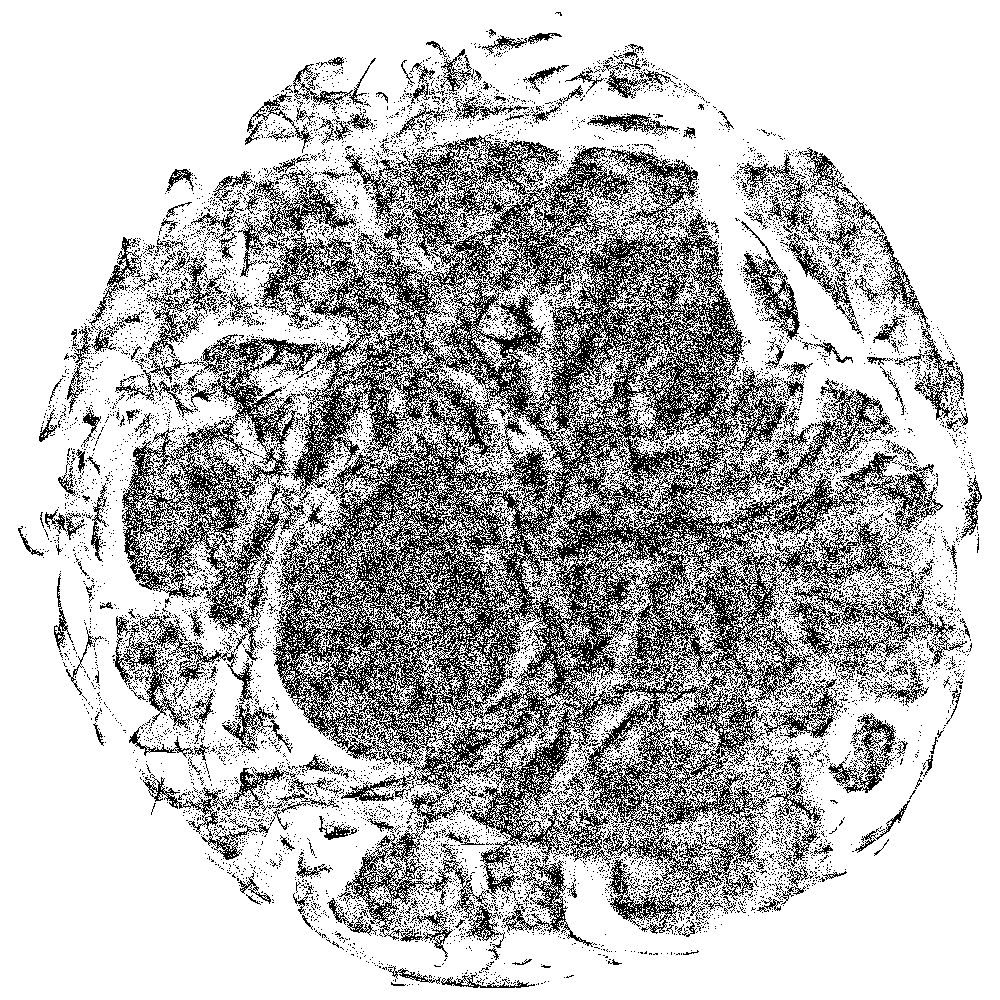}
\includegraphics[width=0.32\textwidth]{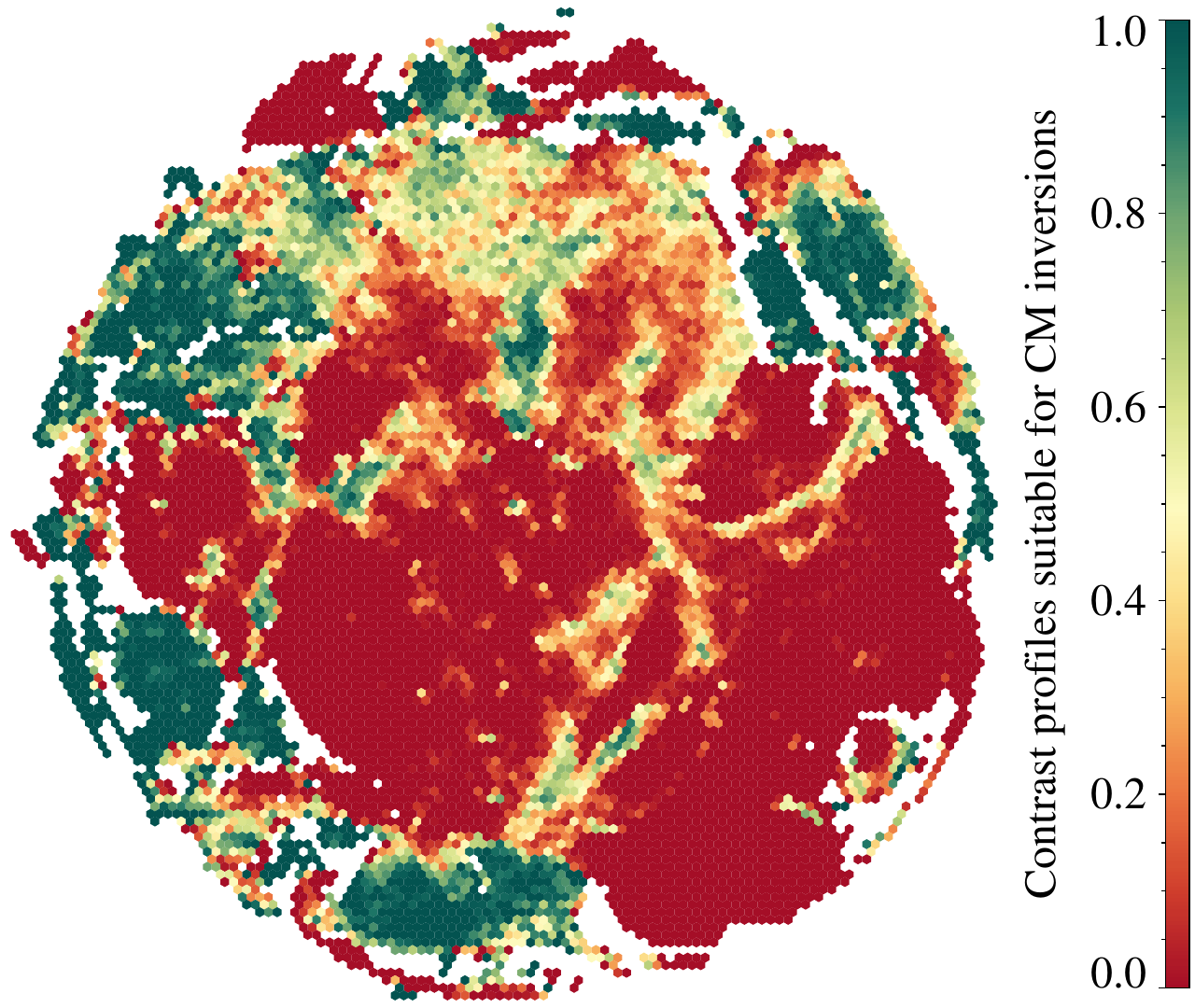}
\includegraphics[width=0.32\textwidth]{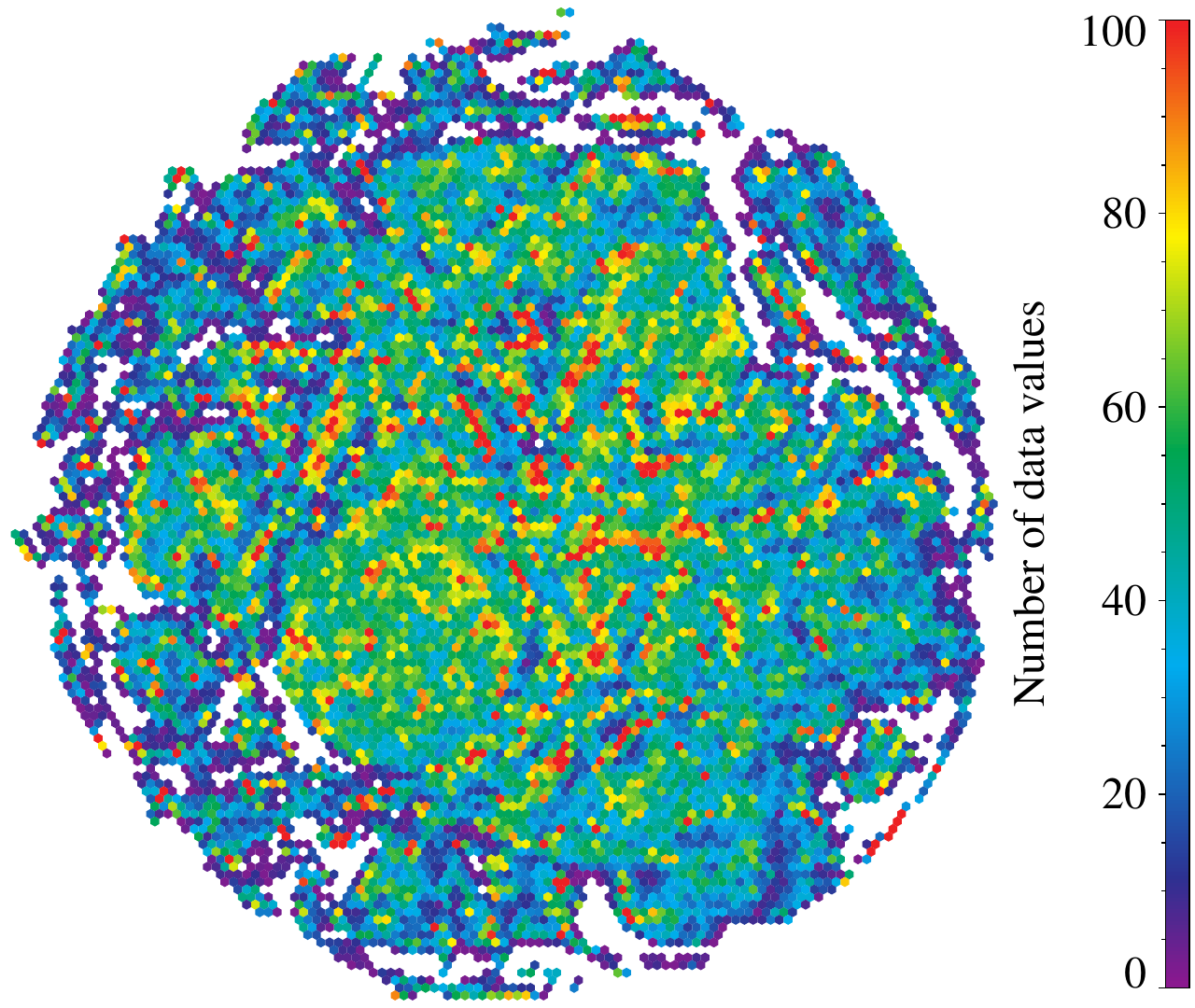}
\caption{Two-dimensional t-SNE projection based on noise-stripped contrast
    profiles, appearing here as a `cloud' of 415\,800 individual points (\textit{left}). Contrast profiles suitable for CM inversions (\textit{green}) are aggregated and projected into two dimensions based on t-SNE classification (\textit{middle}) using hexagonal bins. Quiet-Sun and emission profiles (\textit{red}) show low linear and rank-order correlations when comparing observed and CM contrast profiles. The number of contrast profiles per hexagonal bin (\textit{right}), i.e., a two-dimensional frequency distribution of the projection in the left panel, provides a visual guide to interpret the two-dimensional t-SNE projection.}
\label{FIG02}
\end{figure*}

The following brief description of t-SNE will provide the necessary background for the subsequent data analysis \citep[see][for details]{vanderMaaten2008}. A Gaussian probability distribution centered on each point in this initially 601-dimensional space (corresponding to the wavelength sampling) can be defined with a variance of  ${\sigma_{i}}^2$. The similarity between points $p_i$ and $p_j$ (two profiles) is the conditional probability $P_{j|i}$ for point $p_i$ to pick point $p_j$ as its neighbor. If neighbors were picked in proportion to their probability density defined by the Gaussian distribution, then
\begin{equation}
P_{j|i} = \frac{\exp(-|\!| p_{i} - p_{j}|\!|^2/2 {\sigma_{i}}^2)}
{\sum_{k \neq j} \exp(-|\!| p_{i} - p_{k}|\!|^2/2 {\sigma_{i}}^2)}.\ 
\label{EQ01}
\end{equation}
Since our interest is to find only pairwise similarities, the value of $P_{i|i}$
can be set to zero. For the low-dimensional counterparts $q_i$ and $q_j$ of the 
high-dimensional points $p_i$ and $p_j$, similar conditional probability 
$Q_{i|j}$ can be computed as
\begin{equation}
Q_{j|i} = \frac{\exp(-|\!| q_{i} - q_{j}|\!|^2)}
{\sum_{k \neq j} \exp(-|\!| q_{i} - q_{k}|\!|^2)},\ 
\end{equation}
where the variance is ${\sigma_{i}}^2 = \nicefrac{1}{2}$. This value for ${\sigma_{i}}^2$ is chosen for simplicity as it only results in a rescaled version of the final projection. Since we are again interested in pairwise similarity, $Q_{i|i}$ can also be set to zero. In case the low-dimensional data points $q_i$ and $q_j$ correctly model the high-dimensional data points $p_i$ and $p_j$, then the conditional probabilities $P_{j|i}$ and $Q_{j|i}$ will be equal. However, this never happens in practice. The aim is to arrive at this condition, i.e., to minimize the mismatch between the two distributions. Here, it is obtained by minimizing the sum of the Kullback-Leibler divergence \citep{Kullback1959} over all data points using a gradient descent algorithm. The minimized cost function is given by
\begin{equation}
C = \sum_{i} \sum_{j} P_{i|j}\, \log \frac{P_{i|j}}{Q_{i|j}}.
\end{equation}

Details about the mathematical implementation are presented in \citet{vanderMaaten2008}. In the present study, we used the improved version of the SNE, namely the t-distributed stochastic neighbor embedding (t-SNE). In this version \citep{vanderMaaten2014}, the Barnes-Hut \citep{Barnes1986} algorithm is implemented for faster cost function gradient approximation. We used the multicore implementation of t-SNE.\footnote{\href{https://github.com/biolab/Multicore-TSNE}{github.com/biolab/Multicore-TSNE}} The t-SNE algorithm has a non-convex objective function, which is minimized using a gradient descent optimization. This leads to different solutions for different computation runs, where the results are similar but not exactly the same. Perplexity $p$ and the Barnes-Hut parameter $\theta$ are the two free hyper-parameters, which regulate the generation of the projection. The perplexity $p$ is associated with the number of nearest neighbors that is used in manifold learning algorithms. The variance of the Gaussian distribution in Eq.~\ref{EQ01} is controlled by $p$. The basic understanding is that larger datasets require a larger perplexity value. Moreover, $\theta$ is the parameter, which affects the speed of the Barnes-Hut algorithm. However, its value is a trade-off between speed and accuracy. We discuss these and other parameters in detail in the following section. Using the default values $p = 50$ and $\theta =0.5$, it took about 18~min on a 48-core computer to perform the two-dimensional projection of 415\,800 contrast profiles.

In photospheric images, simple intensity-based thresholding produces often satisfactory classification results, identifying quiet-Sun regions, pores, and sunspots including sub-structures such as penumbra and umbra as well as fine-structures such as penumbral grains and umbral dots. However, the highly structured and dynamic chromosphere is a taxing task for spectral classification, which motivated the present study. The chromospheric region depicted in Fig.~\ref{FIG02a} includes quiet Sun regions, an arch filament system, surges, and some bright plage regions. The main objective of t-SNE is to distinguish among these features in two-dimensional projections of physical parameters. The main challenge is the large number of samples and features, which may obfuscate clear cluster boundaries. Data aggregation as demonstrated in Fig.~\ref{FIG02} benefits visualization but also makes t-SNE results accessible to statistical tools.

The two-dimensional t-SNE projection using contrast profiles of the scan with the best seeing conditions (Fig.~\ref{FIG02}) is based on $n_s = 415\,800$ samples (locations within the FOV) and $n_f = 601$ features (wavelength points). The t-SNE projection of all samples is displayed in the right panel of Fig.~\ref{FIG02}. The $x$- and $y$-axes of the t-SNE projection are machine-learned reduced dimensions, which have no physical significance, and both axes were normalized so that the coordinates are casted in the range $[-1,\, +1]$. The coordinate system is only required to determine distances within and between clusters, whereby the distance should not be taken as a quantitative measure of the classification success. A good classification is characterized by compact, well-separated clusters. However, the large number of samples used for t-SNE projection renders a point-cloud-like appearance of varying density and with several gaps separating center and periphery. In absence of clearly distinct clusters, collecting samples in hexagonal bins and labeling the bins with physical properties (middle panel of Fig.~\ref{FIG02}) are the next steps in classifying spectra. 

The observed chromospheric scene predominantly covers quiet Sun with an embedded activity cluster. The middle panel of Fig.~\ref{FIG02} displays as an example a two-dimensional histogram with about 12\,000 hexagonal bins, where the suitability of contrast profiles for CM inversions is the dependent variable. This is originally a binary parameter, i.e., it is unity if the linear and rank-order correlation coefficients between observed and CM inverted contrast are $\rho_p > 0.95$ and $\rho_s > 0.95$, respectively. In addition, the CM parameters, i.e., optical thickness $\tau_0$, Doppler velocity of the cloud $v_D$, Doppler width of the absorption profile $\Delta\lambda_D$, and source function $S$, have to be within the bounds specified in \citet{Dineva2020}. Only after binning and taking the average, the suitability parameter becomes a floating-point number in the interval $[0.0,\, 1.0]$. Binning and taking the average is carried out using the number of contrast profiles per hexagonal bin as shown in the middle panel of Fig.~\ref{FIG02}.

Particularly, the right panel of Fig.~\ref{FIG02} reveals that the number of the contrast profiles is high in the central hexagonal bins compared to those at the periphery. This is also evident as a higher density of points in the left panel of Fig.~\ref{FIG02}. This arrangement is the most compact representation of a large number of similar quiet-Sun samples. All other profiles with different spectral characteristics are pushed to the periphery, where they form individual clusters. While interpreting the t-SNE maps, this behaviour of the algorithm has to be taken into account.

\begin{figure}[t]
\centering
\includegraphics[width=\columnwidth]{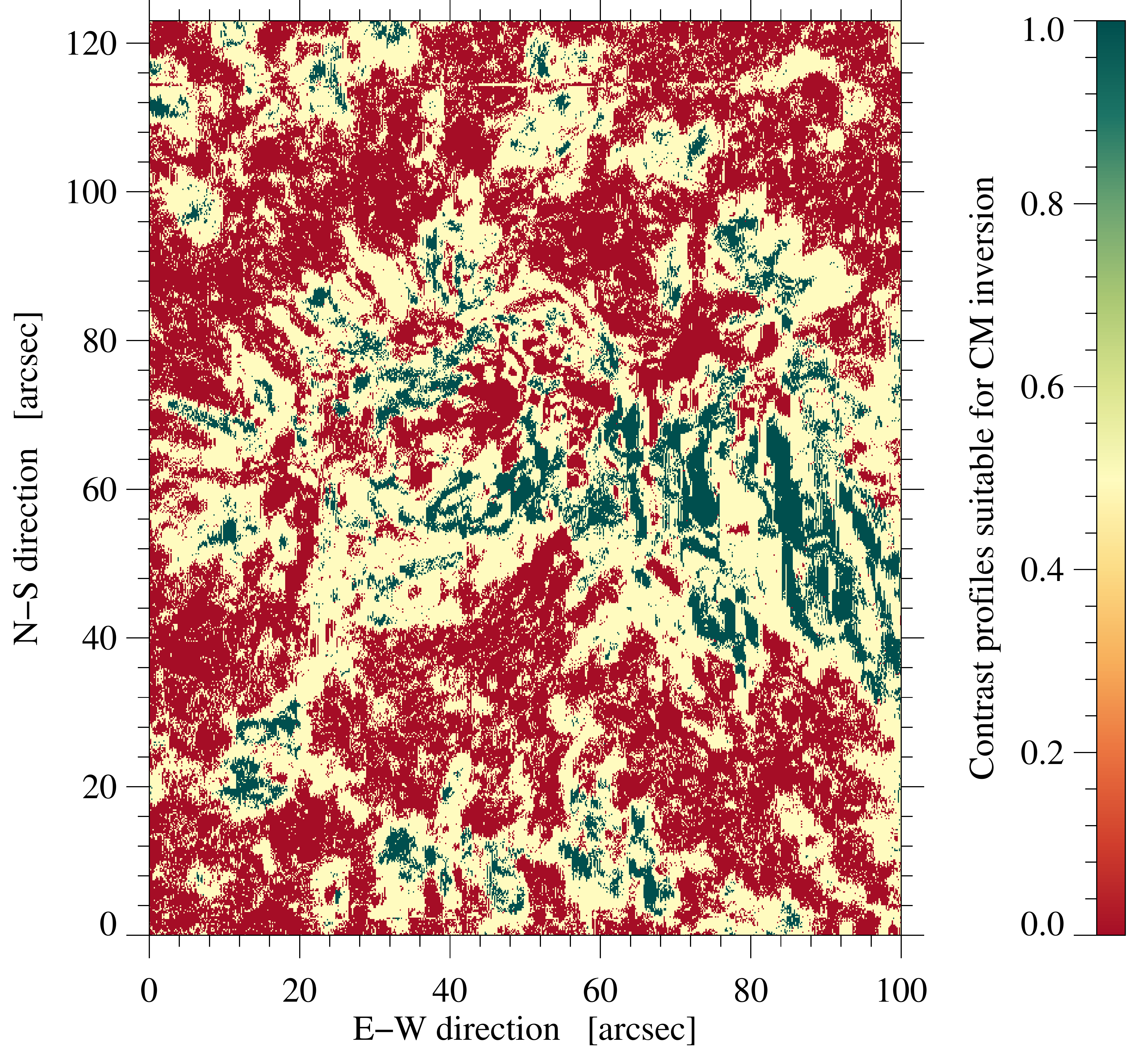}
\caption{Back-projection of contrast profiles suitable for CM inversions
    (\textit{green}) to the observed FOV (same data as in Fig.~\ref{FIG02}). Regions, where CM inversions fail (\textit{red}), represent either the quiet Sun or belong to profiles with enhanced line-core intensities or with emission.}
\label{FIG03}
\end{figure}

In principle, any other physical property of the projected dataset can become the dependent variable with the potential to reveal any clustering in the projection. However, the apparent separation of the red and green colors already indicates two classes of contrast profiles, i.e., those that are suitable for CM inversions (green) and those where CM inversions fail (red). The former profiles belong to dark cloud-like features in H$\alpha$ line-core intensity maps, whereas the latter are associated with quiet-Sun regions and profiles with enhanced line-core intensities or even emission. The identification of the two classes in the t-SNE projection is striking, especially in the absence of any \textit{a priori} knowledge of the underlying physics or implementation of the CM inversion algorithm. This provided the motivation for a detailed parameter study of t-SNE with the goal to find a procedure for the bulk classification of H$\alpha$ spectra, among others, expected from telescopes for high-resolution solar observations.

The output of a t-SNE projection is a two-dimensional coordinate for each contrast profile (left panel of Fig.~\ref{FIG02}), and a one-to-one correspondence exists between these coordinates and those in slit-reconstructed images (Fig.~\ref{FIG02a}). Thus, it becomes possible to back-project physical properties of the t-SNE maps to the observed FOV (Fig.~\ref{FIG03}). This also applies to average values (for example middle panel of Fig.~\ref{FIG02}) and higher stochastic moments (i.e., variance/standard deviation, skewness, and kurtosis) of physical properties, which can be computed for each hexagonal bin (see Sect.~\ref{SEC43}). The interplay between t-SNE map and back projection allows us to associate clusters in t-SNE maps with physical properties of the observed scene on the solar surface. Thus, human inference is still needed to exploit the potential of t-SNE. Figure~\ref{FIG03} illustrates such a back mapping of the data shown in the middle panel of Fig.~\ref{FIG02}. This serves as a sanity check that t-SNE produces meaningful clusters of the input contrast profiles. For clarity of the display, the averaged suitability parameter takes on only three values: zero, unity, and one-half. Compared to the binary map of the suitability parameter, a clearly defined transition zone surrounds the regions with profiles suitable for CM inversions. Within contiguous regions of either unity or zero, other values are rarely encountered and appear as a noise-like pattern. Regions, where the suitability parameter is unity, belong to dark surges, arch filaments, and mottles.

\begin{figure}[t]
\centering
\includegraphics[width=\columnwidth]{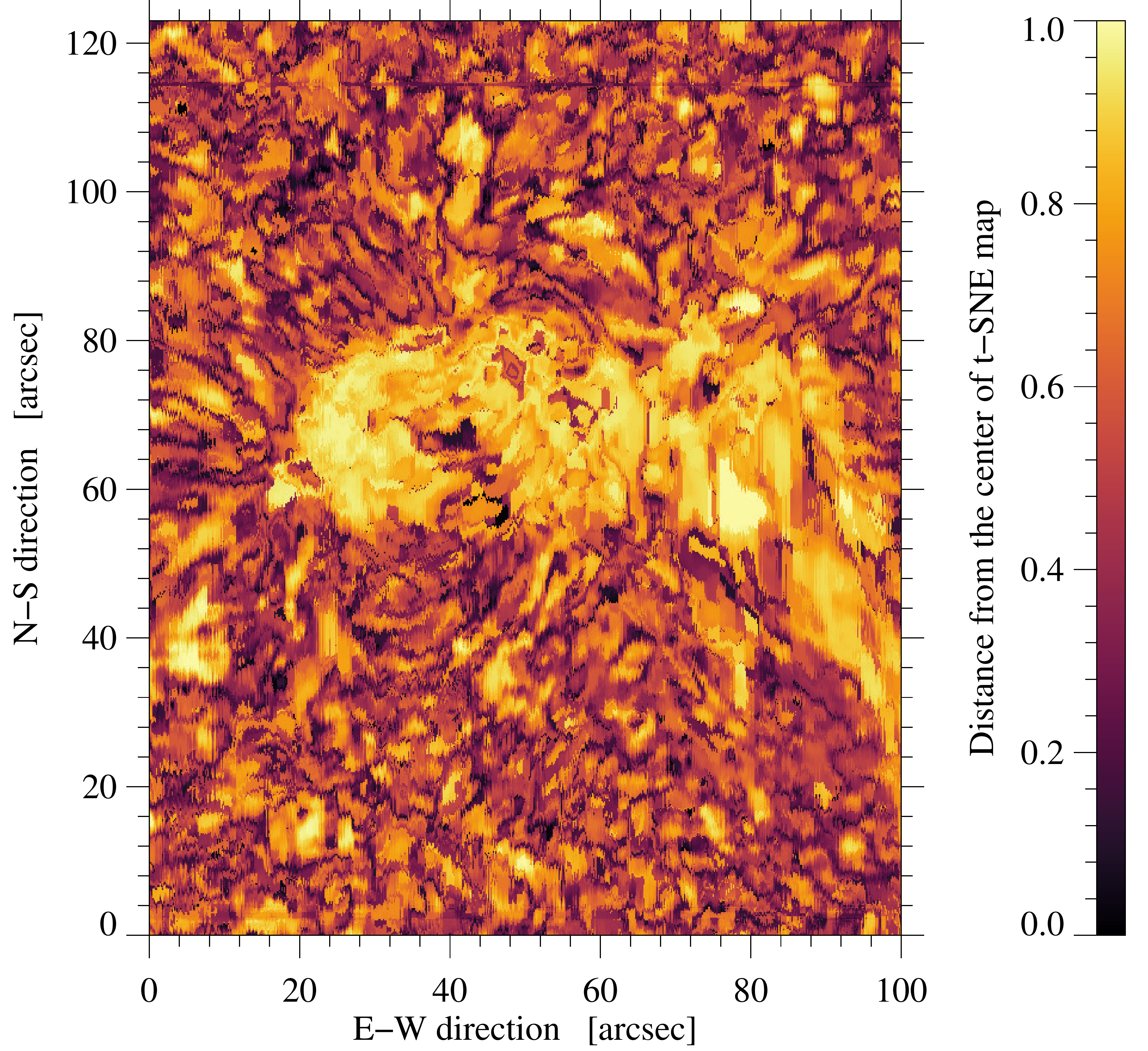}
\caption{Back-projection of the normalized distance from the center
    of the t-SNE projection to the observed FOV (same data as in Fig.~\ref{FIG02}). Zero refers to the center of the t-SNE map, whereas unity indicates the outermost edge.}
\label{FIG04}
\end{figure}

Another way of visualizing t-SNE results is to compute the normalized distance from the center of the t-SNE map and back project it into the observed FOV (Fig.~\ref{FIG04}). The distance from the center of the map is reasonable because the maps tend always to be very circular, almost regardless of how the hyper-parameters are set. The largest distances are found in a compact region in the center of the FOV that encompasses the active region with distinct absorption and emission features in H$\alpha$. This compact region is surrounded by granular-scale clusters of low-distance values, which are typical for the surrounding quiet Sun. Since profiles in quiet-Sun regions outnumber those in active regions, the t-SNE algorithm concentrates the very similar quiet-Sun profiles in the center of the t-SNE map. All other profiles with a broad variety of different shapes are pushed to the periphery. Back projection of the normalized distance is always possible and does not require any additional information beyond the input contrast profiles. However, a clear separation as in Fig.~\ref{FIG04} will not always occur and depends on the observed features and their frequency of occurrence within the FOV.


\section{Results} \label{sec:results}

In this section, we present a parameter study to find an optimal t-SNE setup adapted to input data, i.e., contrast profiles of the chromospheric H$\alpha$ line. Once the optimal setup is established, we explore back-mapping and re-projection of selected clusters in the t-SNE maps. Furthermore, classification of profiles based on t-SNE projection are presented and discussed.


\subsection{Parameter study}\label{SEC41}

Frequently, t-SNE projections are used for data dimensionality reduction and visualization. However, to interpret and understand the projection, the user's expertise and domain knowledge of the input data are needed. In addition, an acute awareness of the parameters that control the t-SNE projection of the data is required, in order to avoid misinterpreting the results.
Therefore, it is important to grasp how the t-SNE projection depends on perplexity $p$, Barnes-Hut parameter $\theta$, and number of iterations $n$. The multicore implementation of t-SNE depends only on these parameters, which are investigated in the following parameter study. A detailed account of these parameters is given by \citet{Wattenberg2016} who presented a graphical description of t-SNE using various combinations of parameters.

The default parameters of t-SNE are Barnes-Hut parameter $\theta=0.5$, perplexity $p=50$, and number of iterations $n = 1000$. Meaningful parameters, i.e., parameters that were commonly used in other studies, cover the intervals $\theta \in [0.2,\, 0.8]$, $p \in [10,\, 100]$, and $n \in [200,\, 4000]$ resulting in numerous triples of parameters. For example, the selected parameter range of the perplexity is based on \citet{vanderMaaten2008}, who recommended a typical range of 5\,--\,50. However, the three-dimensional space spanned by the t-SNE parameters can be truncated to save computing time while preserving the most notable dependencies. Two parameters are kept fixed at the default values while changing the third one. This scheme produces 13 t-SNE maps and allows us a comprehensive comparison of the t-SNE projections (Fig.~\ref{FIG03}). The input data are again the contrast profiles of the best scan, and the maps are color-coded as before to discern if the observed contrast profiles are suitable for CM inversions. 

\begin{figure*}
\centering
\includegraphics[width=\textwidth]{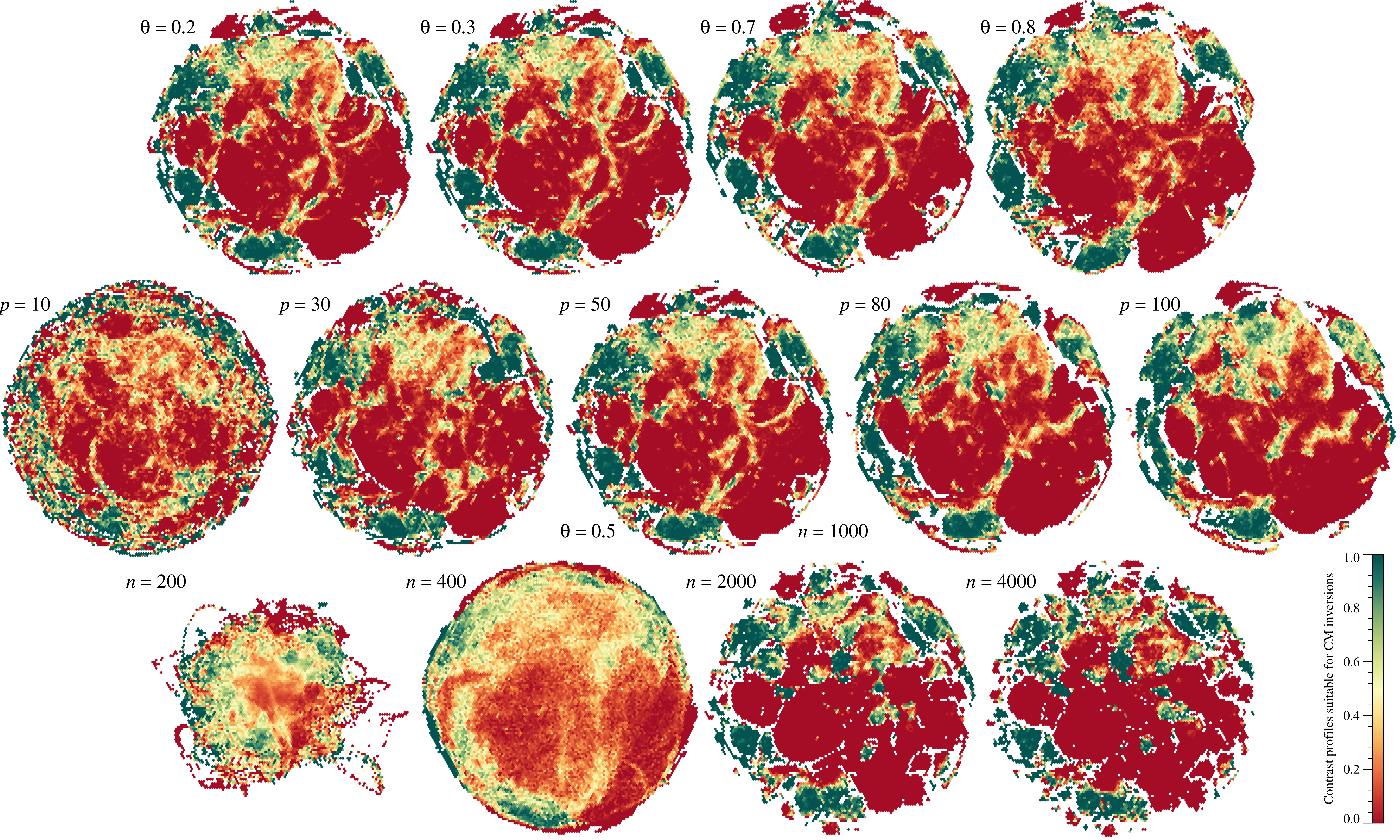}
\caption{Parameter study of t-SNE projections as a function of $\theta$ 
    (\textit{top row}), perplexity $p$ (\textit{middle row}), and number of iterations $n$ (\textit{bottom row}). The projection with $\theta=0.5$,  perplexity $p=50$, and number of iterations $n=1000$ refers to the defaults settings (\textit{third panel, middle row}). The two-dimensional projections are color-coded in the same way as in the middle panel of Fig.~\ref{FIG02} and in Fig.~\ref{FIG03}.}
\label{FIG05}
\end{figure*}

The top row in Fig.~\ref{FIG05} demonstrates that varying the Barnes-Hut parameter $\theta$ only leads to minute changes in the t-SNE maps, which are mainly restricted to the periphery of the maps, in particular for high values of $\theta$. Changing the perplexity $p$ from low to high values (middle row of Fig.~\ref{FIG05}), profiles that are not suitable for CM inversions become increasingly concentrated in the center of the map. The values in the periphery of the t-SNE map are initially almost randomly distributed but they become more clustered for $p>30$, in particular for the profiles that are suitable for CM inversions. Once $p=50$ is reached, the difference between the maps are again minute. If the number of iterations is very low ($n=200$ in the bottom row of Fig.~\ref{FIG05}), the circular coordinate space of the t-SNE map is only incompletely filled, which changes to the contrary for $n=400$. In the latter case, the hexagonal bins cover the entire disk-shaped region but predominant yellow and orange colors indicate a deficient classification of the contrast profiles. A clear clustering of the t-SNE maps becomes apparent at $n = 1000$, i.e., a large number of iteration is needed for large datasets to optimize the t-SNE projection. Increasing the number of iterations further does not change the morphology of the maps and only slightly boosts the quality of the classification -- at the cost of rising computational effort. The computing time using 48 of 64 cores of a compute server (AMD Opteron 6378) takes 18, 32, and 63~min for $n = 1000$, 2000, and 4000, respectively. Thus, the computing time depends almost linearly on the number of iterations $n$.

\begin{figure*}
\centering
\includegraphics[width=\textwidth]{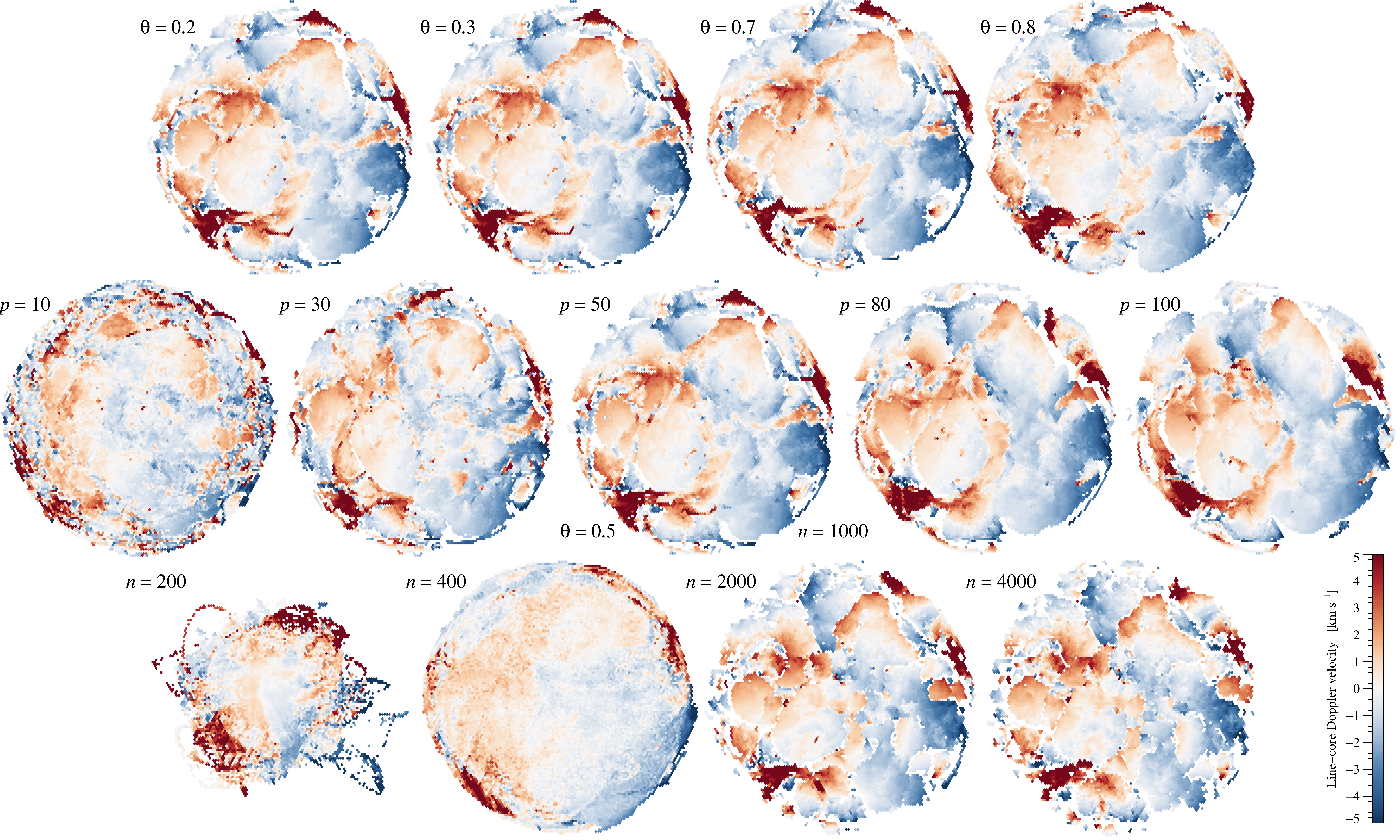}
\caption{Parameter study of t-SNE projections as a function of $\theta$ 
    (\textit{top row}), perplexity $p$ (\textit{middle row}), and number of iterations $n$ (\textit{bottom row}). The projection with $\theta=0.5$,  perplexity $p=50$, and number of iterations $n=1000$ refers to the defaults settings (\textit{third panel, middle row}). The two-dimensional projections are color-coded according to the line-core Doppler velocity.}
\label{FIG06}
\end{figure*}

Line width and Doppler shift of spectral lines are the most obvious parameters affecting morphology and position of spectra. Therefore, we expect that they have a significant impact on the classification. Using the line-core Doppler shift, which was derived from parabola fitting of the line core, the 13 t-SNE maps are plotted in Fig.~\ref{FIG06} with the same layout and parameters sets as in Fig.~\ref{FIG05}. Negative and positive plasma flows, i.e., blue- and redshifts of the spectral line form patterns that are aligned with clusters that are already apparent in Fig.~\ref{FIG03}, regardless if these clusters refer to contrast profiles that are suitable or unfit for CM inversions. Exceptions are encountered for low values of the perplexity ($p=10$ and $p=30$) and small numbers of iterations ($n=200$ and $n=400$). The morphology of the remaining 11 maps is very similar. The Doppler velocity serves as a secondary criterion to separate the flow speed within each cluster. Thus, sharp transitions from positive to negative velocities provide clues where the borders of clusters may be located. In addition, the highest flow speeds are encountered at the periphery of the t-SNE maps because they are by definition absent in quiet-Sun regions at the center of the maps. Therefore, the t-SNE perform correctly the coarse classification between profiles characteristic for the active and quiet Sun without \textit{a priori} knowledge of the underlying physics. In summary, comparing the t-SNE projections and computing time for various combinations of parameters, we conclude that the default parameters $\theta = 0.5$, $p = 50$, and $n = 1000$ are already a very good choice for our dataset.

\subsection{Choice of input data and impact of seeing conditions}

\begin{figure*}
\centering
\includegraphics[width=\textwidth]{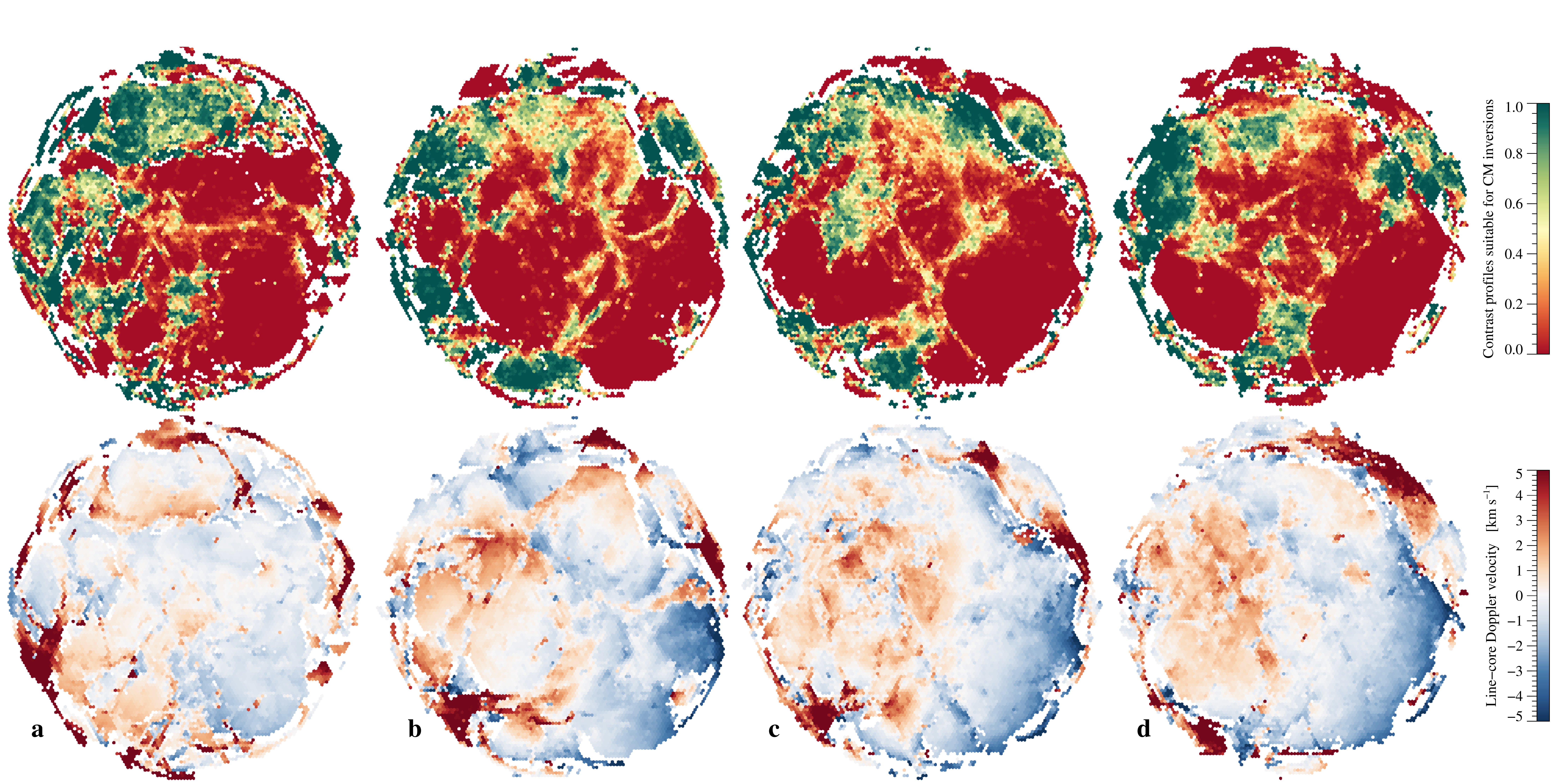}
\caption{Comparison of t-SNE projections using different input data: (a)
    noise-stripped contrast profiles for a bad seeing scan, (b) PCA coefficients for the best seeing scan, (c) observed contrast profiles for the best seeing scan, and (d) noise-stripped intensity profiles for the best seeing scan. All projections are color-coded according to their suitability for CM inversions (\textit{top}) and depending on the line-core Doppler velocity (\textit{bottom}).}
\label{FIG07}
\end{figure*}

Apart from optimizing the hyper-parameters $\theta$, $p$, and $n$ of the t-SNE projection, various forms of input data will result in different t-SNE projections, which are scrutinized in this section. Ground-based observations are affected by varying seeing quality. This raises the question whether seeing conditions impact t-SNE projections. In addition, input data can be contrast or intensity profiles, either observed or noise-stripped using PCA, or just the PCA coefficients of the first 10 eigenfunctions. We investigated these issues by creating t-SNE projections for some variants of input data, which are compiled in Fig.~\ref{FIG07}.
\newpage

As noted in Sect.~\ref{sec:methods}, different computational runs will result in somewhat different t-SNE projections. Hence, it is impossible to carry out a one-to-one comparison of the t-SNE projections in Fig.~\ref{FIG07}. Yet, the overall morphology provides some guidance regarding quality and choice of input data. Red and green values are well separated and the profiles suitable for CM inversions are in general pushed to the periphery. However, the maps for bad seeing data appear noisier (e.g., scan No.~20). In particular, regions are not as compact, where the contrast profiles are unfit for CM inversions. Interestingly, the t-SNE maps based on just the PCA coefficients (Fig.~\ref{FIG07}b) are virtually identical to the reference dataset (middle panel of Fig.~\ref{FIG02} and central maps in Figs.~\ref{FIG05} and~\ref{FIG06}), indicating that 10 eigenfunctions are sufficient to capture the essential morphology of H$\alpha$ contrast profiles. There are subtle indications (darker green clusters in Fig.~\ref{FIG07}b as compared to Fig.~\ref{FIG07}c) that PCA-based noise-stripping performs better in identifying contrast profiles for CM inversions. A similar trend is evident for the t-SNE projection using noise-stripped intensity profiles (Fig.~\ref{FIG07}d). Thus, noise-stripping is a distinct advantage for classifying H$\alpha$ profiles of chromospheric absorption features. Since the average computing time (about 18~min) was the same for all t-SNE maps, it is not an important criterion for choosing the input data for t-SNE.

\subsection{t-SNE projections of CM inversions}\label{SEC43}

Much of the chromospheric complexity of the active Sun is contained in absorption profiles of cool plasma compared to its surroundings that is suspended by the magnetic field above the solar surface. Before exploiting the results from Sects.~\ref{sec:methods} and~\ref{SEC41}, we apply t-SNE to all H$\alpha$ contrast profiles of the best scan, which fulfill the criteria $\rho_p > 0.95$ and $\rho_s > 0.95$, i.e., they produce reliable CM inversion. The results are compiled in Fig.~\ref{FIG08}, which presents t-SNE projections for the four CM parameters, i.e., Doppler velocity of the cloud $v$, optical depth $\tau$, source function $S$, and Doppler width $\Delta\lambda_D$. 

Visually, the t-SNE maps of the CM parameters can be characterized by their complexity. The map of the Doppler width $\Delta\lambda_D$ is the simplest, where a vertical line is sufficient to roughly separate small and large values. The maps of optical depth $\tau$ and source function $S$ both display a wedge-like intrusion from the bottom. The map of the source function $S$ basically consists of three regions, i.e., a large uniform region of low values to the right (light green), a smaller region with medium values to the left (medium blue-green), and the wedge-like intrusion from the bottom (dark blue). Overall, the map of optical depth $\tau$ has an almost inverse look compared to the source function $S$. However, here the wedge-like region is more uniform whereas the other regions reveal gradients. Morphological image processing could be applied to the maps of optical depth $\tau$ and source function $S$ to clearly separate the three classes. Finally, the map of the cloud velocity $v$ possesses the highest level of complexity. However, on closer inspection, it is just the velocity gradient that further differentiates the aforementioned three classes. In summary, visualizing patterns or clusters in t-SNE maps is a powerful tool of human inference, enabling a physics-based classification of H$\alpha$ contrast profiles.

We plot the standard deviation, which represents the second central moment of the probability distribution for each hexagonal bin, for the four CM parameters in Fig.~\ref{FIG10a}. In standard deviation maps of Doppler width $\Delta\lambda_D$, optical depth $\tau$, and source function $S$ a wedge-like intrusion from the bottom is evident. In the case of source function $S$ and Doppler width $\Delta\lambda_D$, this belongs to large values of standard deviation, whereas for optical depth $\tau$ this trend is inverse. For cloud velocity $v$ the appearance of standard deviation map is as complex as the original map. In general, the large values of standard deviation for all four CM parameters belong to the regions with large values in the maps compiled in Fig.~\ref{FIG08}.


\subsection{Classification of CM-invertible contrast profiles}\label{SEC44}

To go one step further, we exploit the clustering visible in the t-SNE projection.
These distinct islands are easily recognizable in t-SNE projection and comprise spectral profiles with similar shape. We identify the ten largest contiguous clusters (left panel of Fig.~\ref{FIG09}), where the hexagonal bins exceed the threshold of 0.9 for the suitability parameter (see Sect.~\ref{SEC43}). This is achieved by applying a threshold of 0.9 to the two-dimensional histogram with hexagonal bins. In the next step, contrast profiles have to be labeled, which belong to the hexagonal bins above the threshold. The relatively high threshold preserves some of the binary character of the suitability parameter (see Sect.~\ref{sec:methods}) while adding clustering information from the t-SNE algorithm. The ten largest, contiguous clusters were selected from the periphery of the t-SNE projection. They are color-coded and numbered in clock-wise direction. Four out of ten cluster are isolated while the remaining one appears in pairs (2\,\&\,3, 4\,\&\,5, and 8\,\&\,9), which are separated by narrow lanes. Since the information about the location of the profiles is available, the cluster can be back-projected to the two-dimensional slit-reconstructed map of, for example, the H$\alpha$ line-core intensity (middle panel of Fig.~\ref{FIG09}). 

\begin{figure*}
\centering
\includegraphics[width=0.24\textwidth]{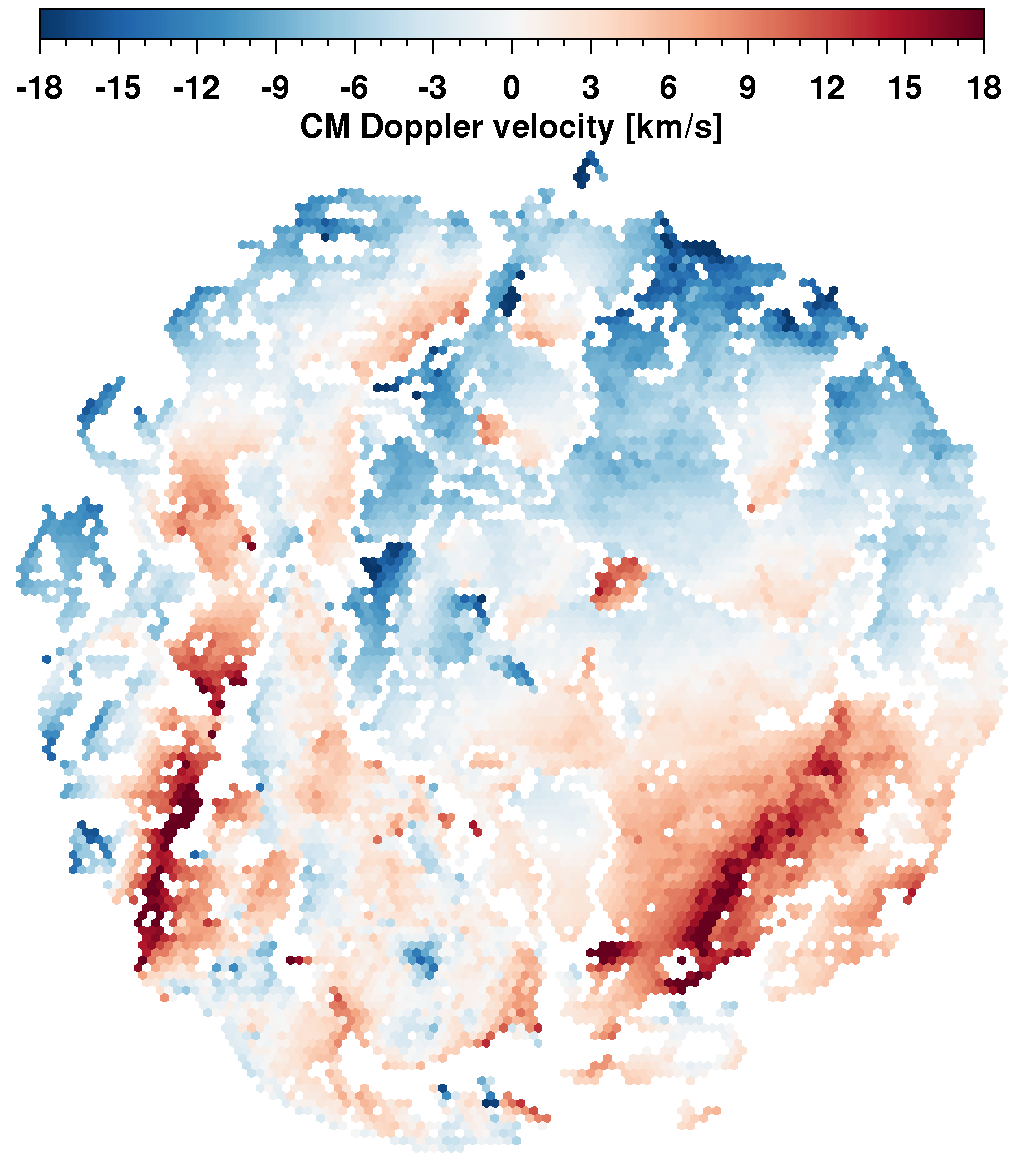}
\includegraphics[width=0.24\textwidth]{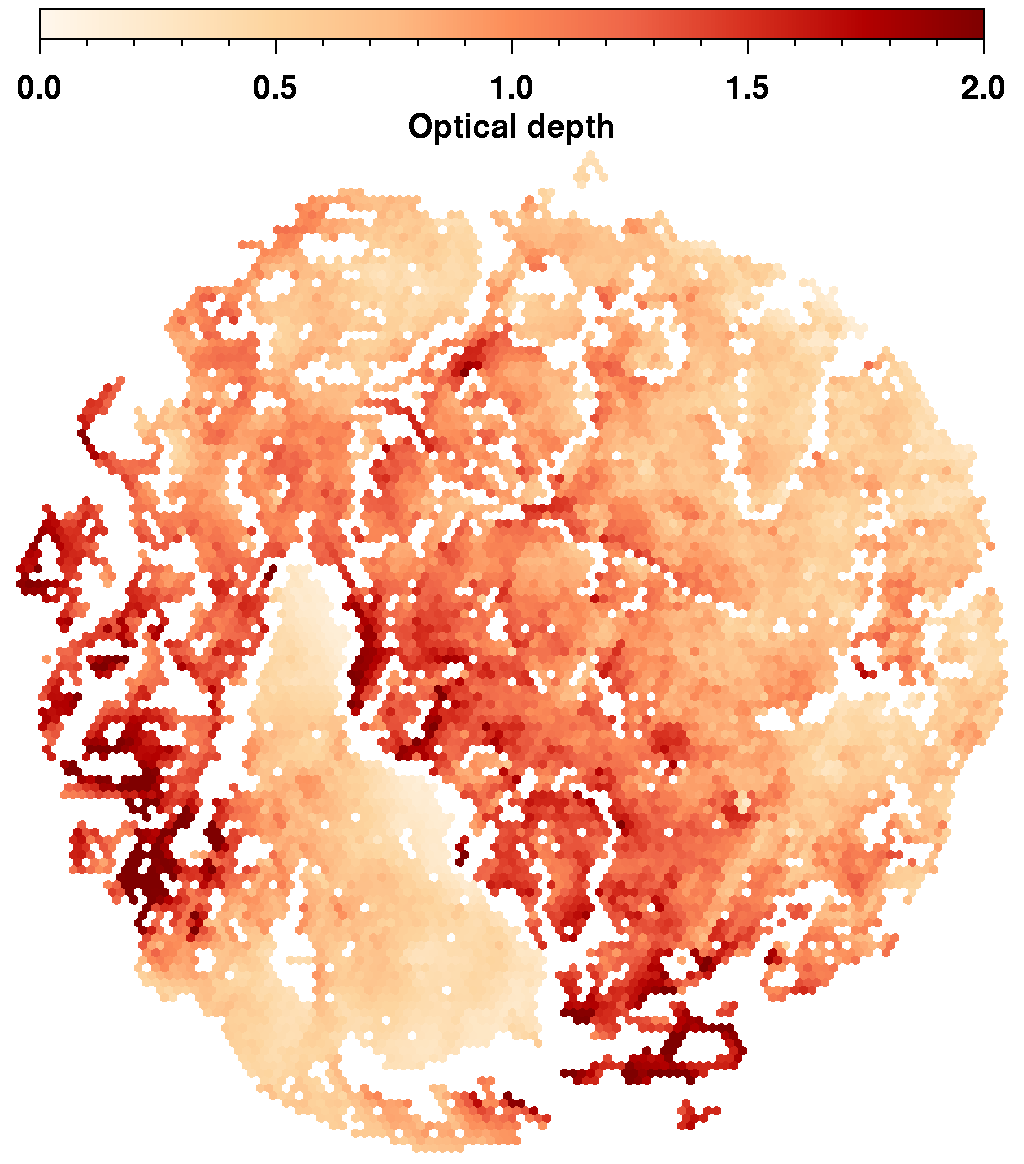}
\includegraphics[width=0.24\textwidth]{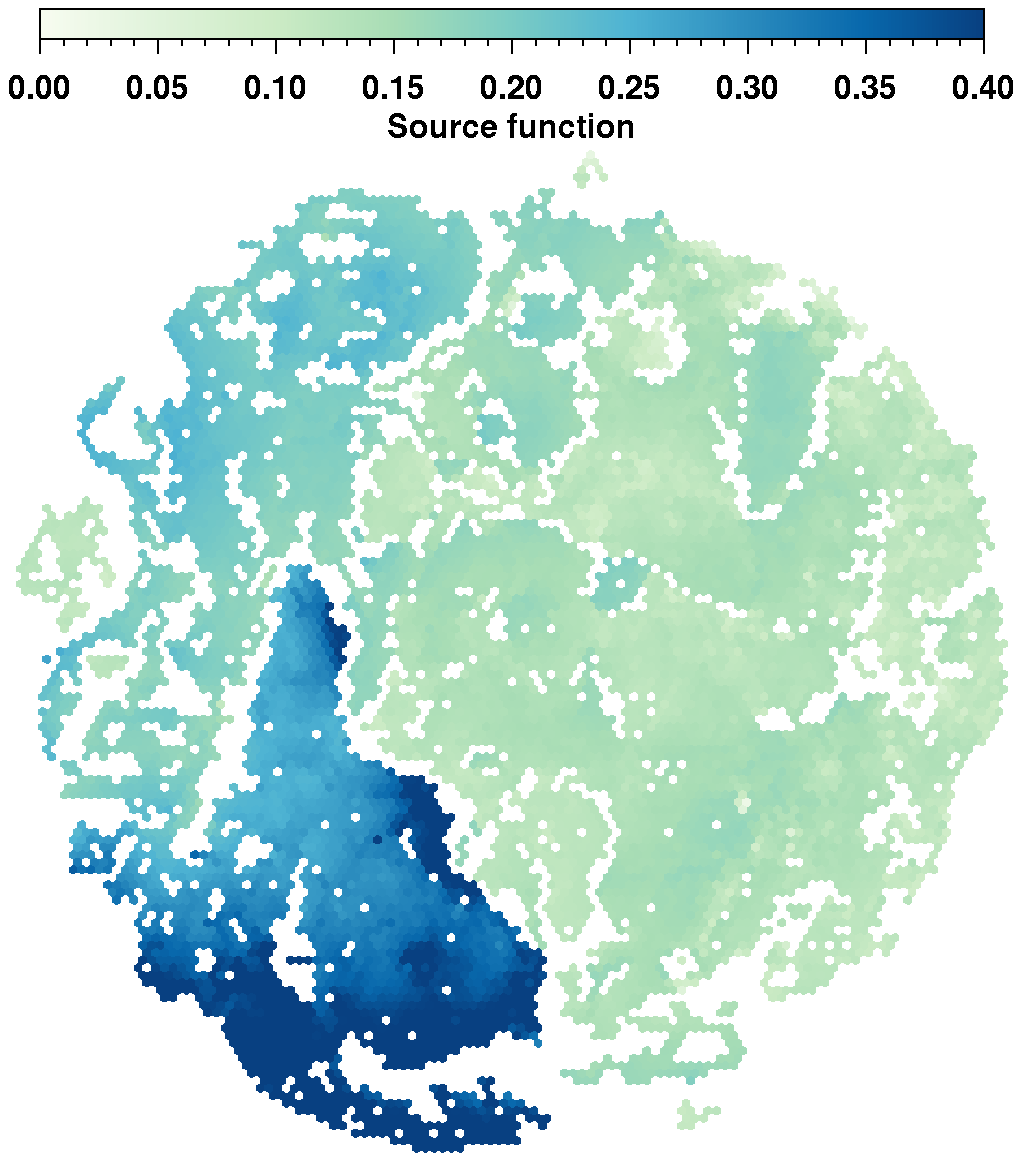}
\includegraphics[width=0.24\textwidth]{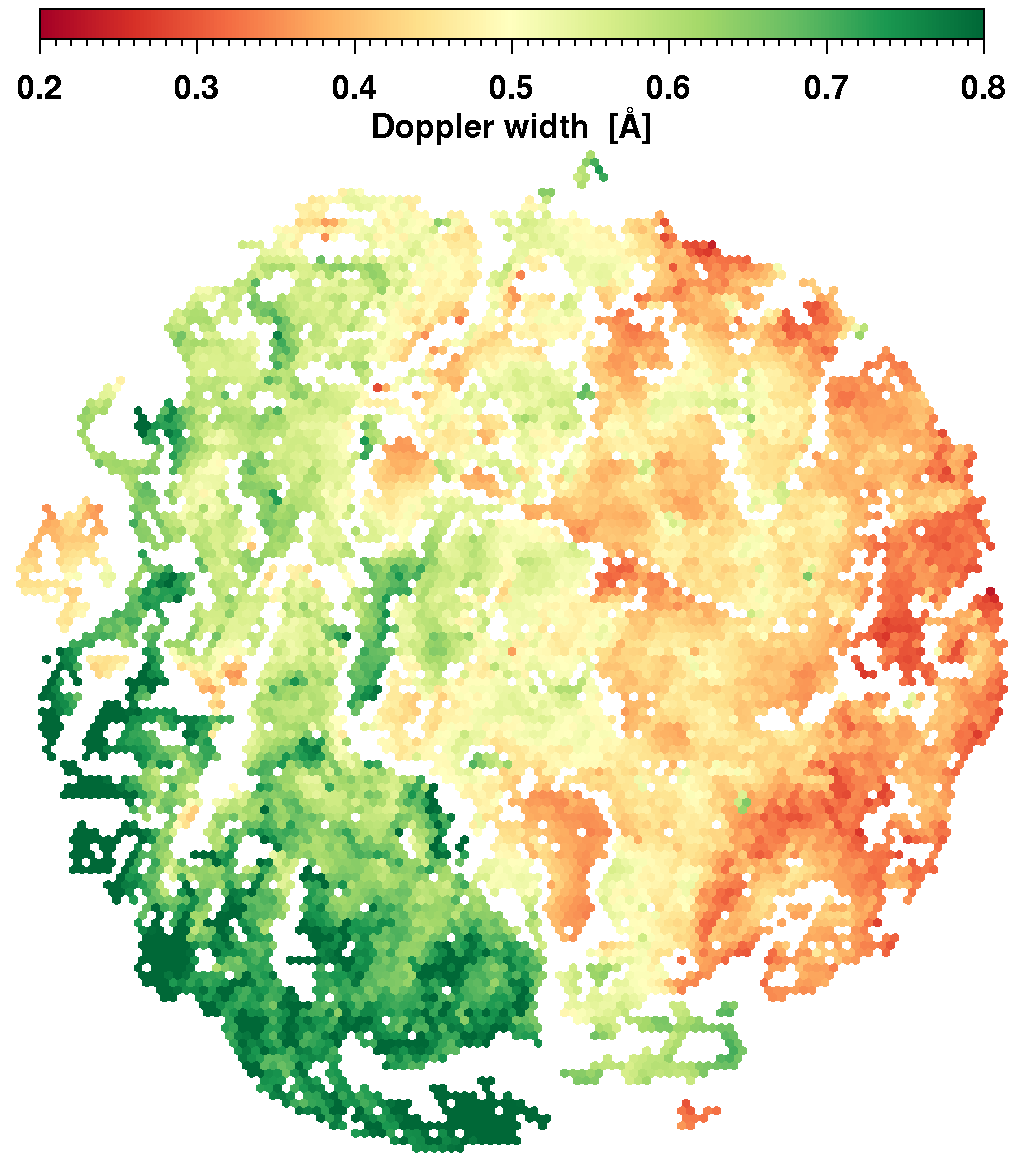}
\caption{Contrast profiles, which are suitable for CM inversions, are used for
    t-SNE projections of the four CM parameters: cloud velocity $v$, optical depth $\tau$, source function $S$, and Doppler width $\Delta\lambda_D$ (\textit{from left to right}). The scale bars at the top of each panel correspond to the range of the CM parameters.}
\label{FIG08}
\end{figure*}

\begin{figure*}
\centering
\includegraphics[width=0.24\textwidth]{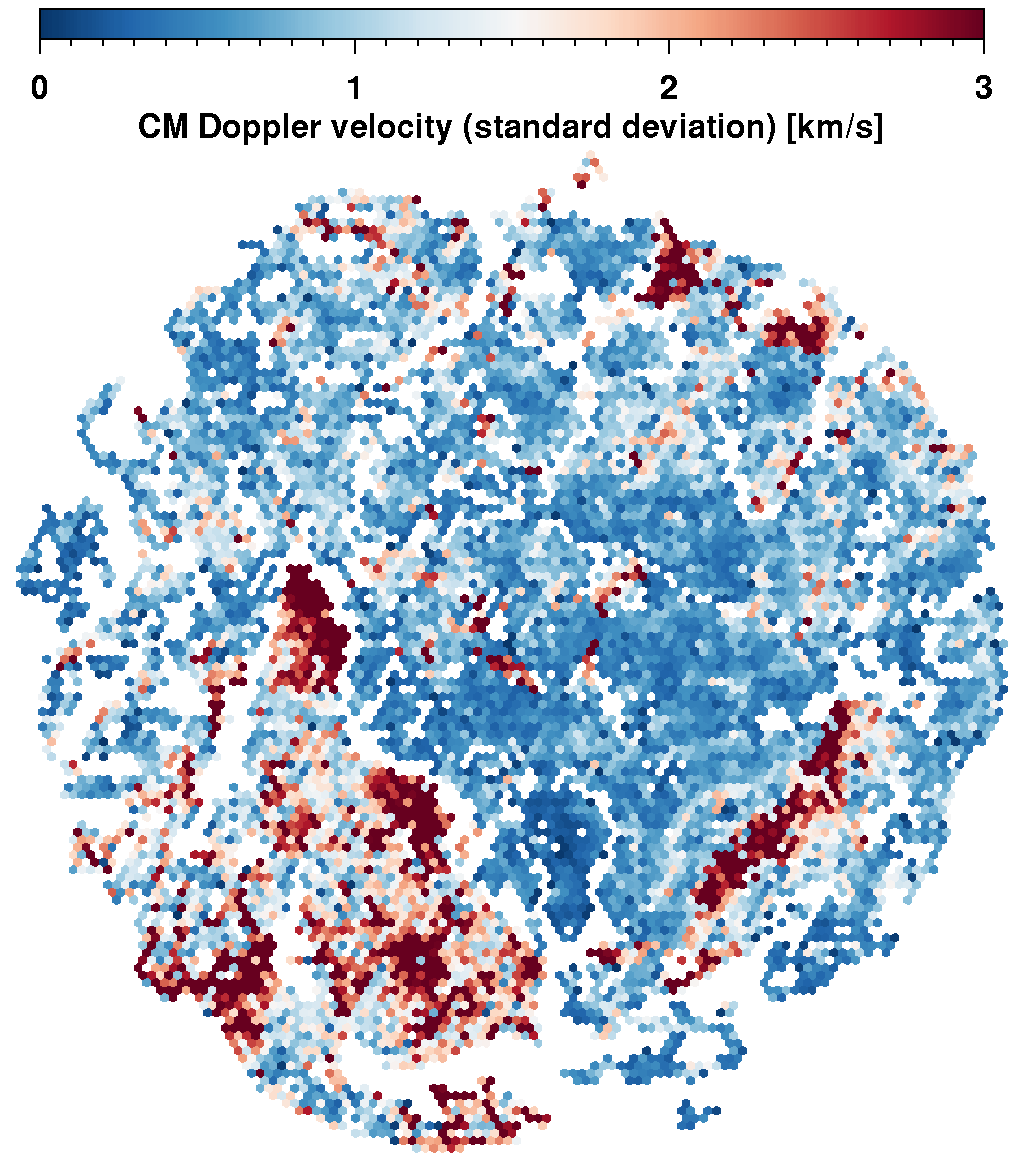}
\includegraphics[width=0.24\textwidth]{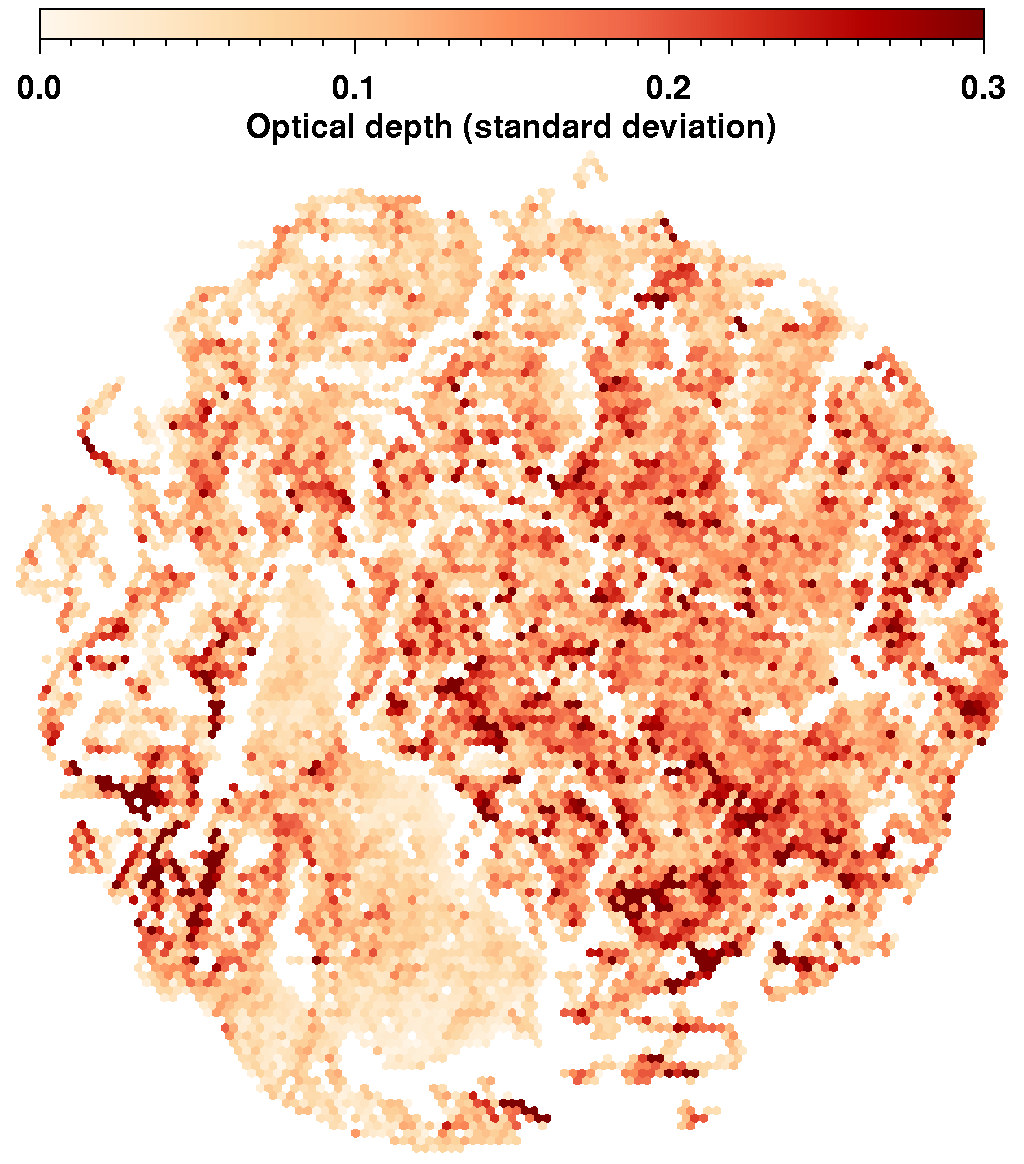}
\includegraphics[width=0.24\textwidth]{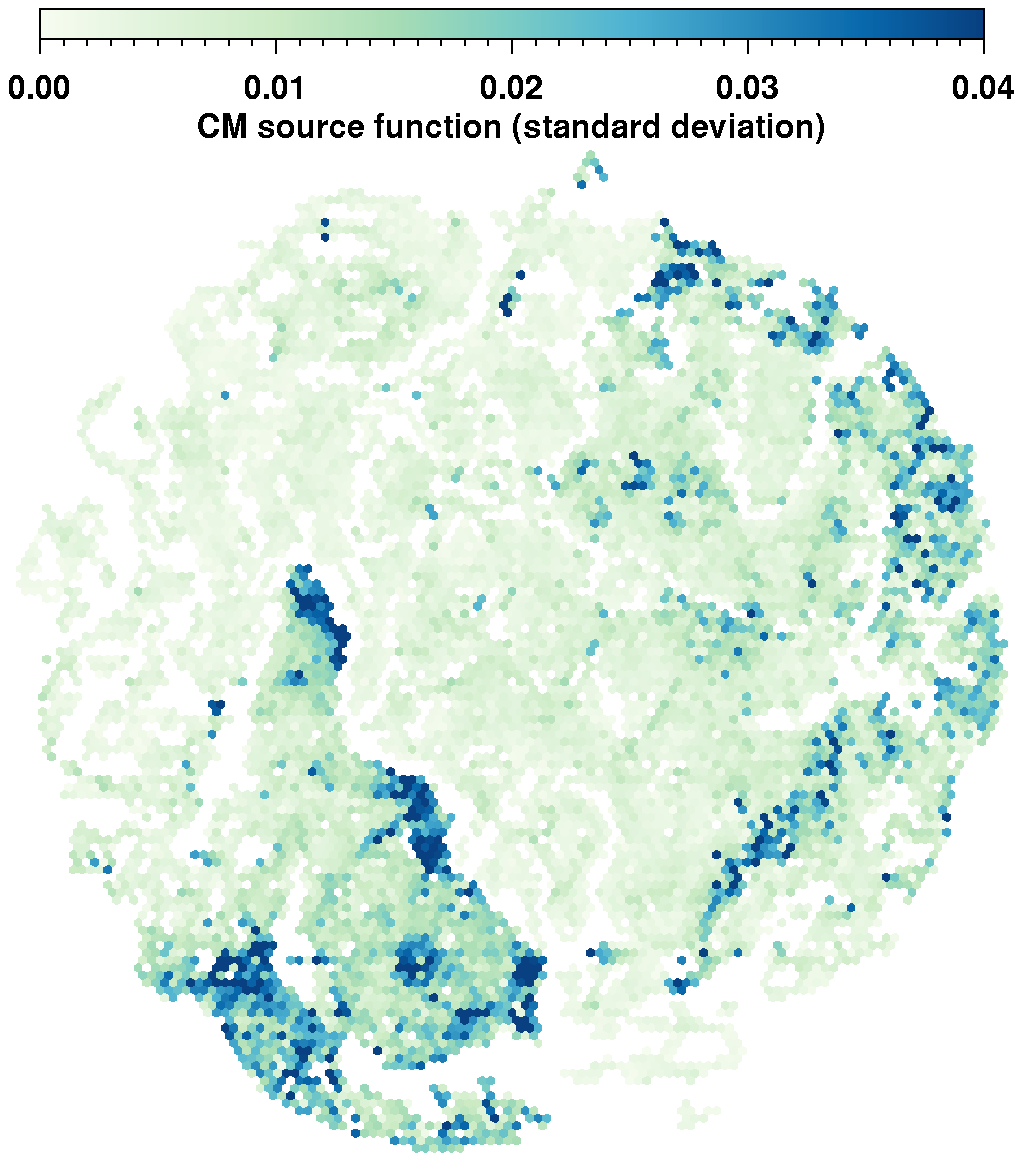}
\includegraphics[width=0.24\textwidth]{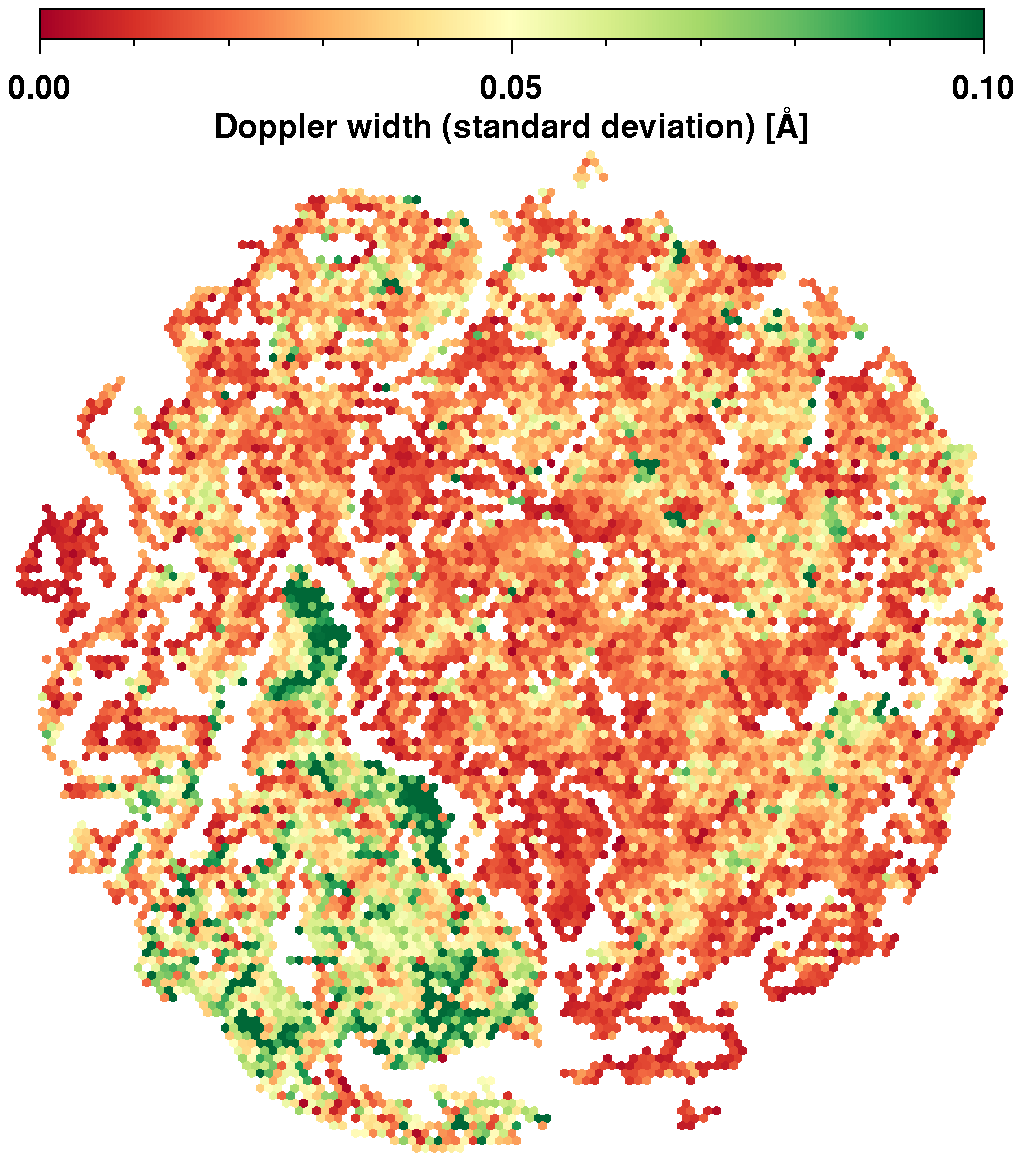}
\caption{Standard deviation of the four CM parameters, i.e., cloud velocity $v$, optical depth $\tau$, source function $S$, and Doppler width $\Delta\lambda_D$ (\textit{from left to right}), corresponding to t-SNE projections shown in Fig.~\ref{FIG08}.}
\label{FIG10a}
\end{figure*}

The result of this back-mapping is compiled in the middle panel of Fig.~\ref{FIG09}, which establishes the link between clusters and different chromospheric absorption features. Almost all clusters are associated with the dark, cloud-like plasma encountered predominantly in the surge and arch filaments but also in regions of mottles surrounding the active region. Cluster~1 profiles mainly refer to arch filaments and the outer edge of the surge. Cluster~2 profiles are associated with footpoint regions of surges and darker regions of arch filaments. Clusters~4 and~5 are related to the upper and middle region of the surge, respectively. A small patch associated with Cluster~4 constitutes a dark region in H$\alpha$ line-core map located away from the central FOV. Clusters~6, 7, 8, and~9 are scattered and not clearly linked to specific features. Cluster~10 contains a set of profiles with enhanced line-core intensity near the base of surge as well as near the footpoints of arch filaments. So far we only referred to clusters as opposed to classes because the term class should be reserved to H$\alpha$ contrast or intensity profiles with the same or at least similar physical origin.

As a sanity check, we subjected the thresholded and clustered profiles to an additional round of t-SNE to determine if the clusters and their spatial relationship survive. The results are depicted in the right panel of Fig.~\ref{FIG09}. By and large, the initial clusters are also present in the new projection. However, the pairwise relationship remained only for Clusters~4\,\&\,5, whereas a new pair was formed for Clusters~1\,\&\,3. Once more, human inference is required in t-SNE to establish meaningful physical relationships among clustered profiles.

Therefore, we compiled in Fig.~\ref{FIG10}, three hundred randomly selected intensity and contrast profiles for each of the ten clusters. In addition, average contrast profiles for each cluster are also displayed to identify characteristic profiles shapes. Similarities and differences in the profiles are clearly evident for the ten selected clusters. Profiles belonging to Cluster~1 exhibit enhanced line-core intensities and are shifted to blue wavelengths indicating plasma upflows. Profiles are broad in Clusters~2, 3, and~5 and have line-core intensities close to those of the quiet Sun. They display blue-shifted asymmetries, which become stronger from Cluster~3, over~5, to~2. Cluster~4 contains deep, unshifted profiles while the deep profiles of Clusters~6, 8, and~7 (ordered according to decreasing line depth) extend mostly to the blue. Profiles in Clusters~9 and~10 have significantly enhanced line-core intensities, whereby the profiles belonging to Cluster~9 are broadened and unshifted while those of Cluster~10 are extremely broad and asymmetric, including well pronounced shoulders indicative of dual-flow components. A similar behaviour is observed in the contrast profiles, where the amplitude variations facilitate an easier recognition of spectral characteristics.

\begin{figure*}
\centering
\includegraphics[width=\textwidth]{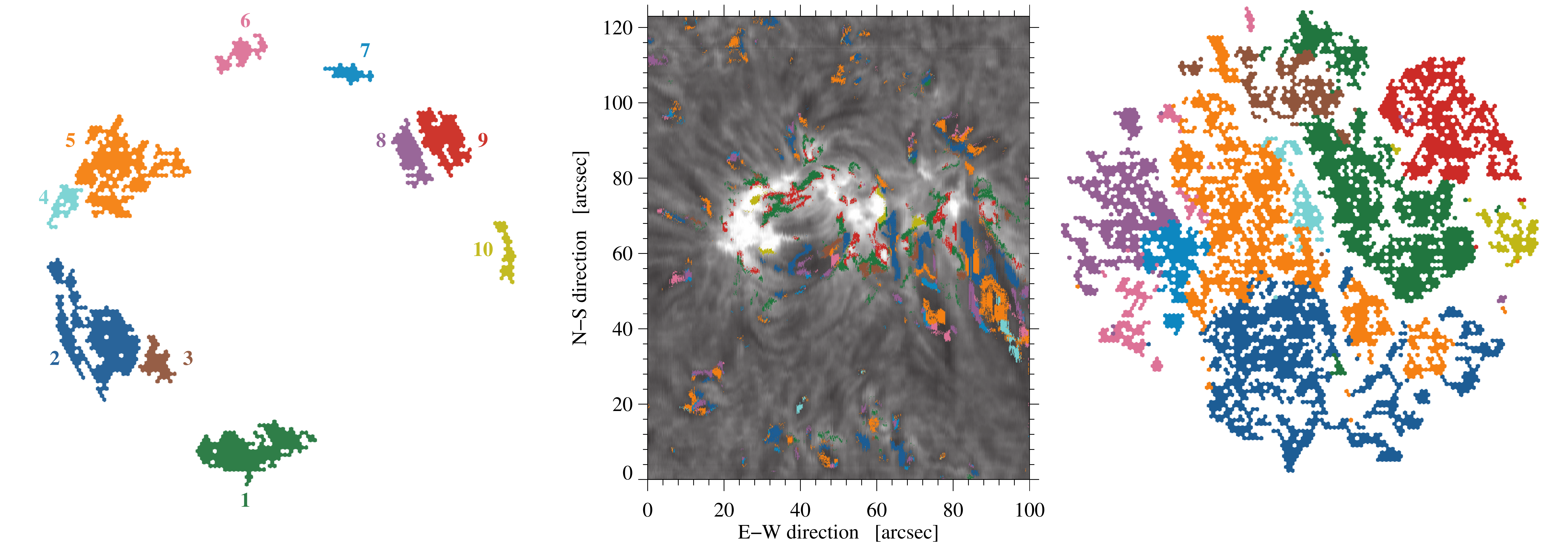}
\caption{Clusters of H$\alpha$ contrast profiles (\textit{left}), which are
    suitable for CM inversions and which belong to the ten largest clusters in the two-dimensional histogram with hexagonal bins. The clusters are depicted in different colors and labeled by numbers. Back-projection of the ten clusters to the slit-reconstructed H$\alpha$ line-core intensity image (\textit{middle}) reveals their relationship to absorption features in the active region. Limiting the t-SNE input data to the back-projected contrast profiles yields a t-SNE projection that clearly preserves cluster membership (\textit{right}).}
\label{FIG09}
\end{figure*}

As previously discussed, some clusters may form classes. Combining the re-projected cluster in the right panel of \mbox{Fig.~\ref{FIG09}} with the spectral characteristics summarized in \mbox{Fig.~\ref{FIG10}} yields three classes: (1) Contrast profiles with a pronounced central component, i.e., Clusters~3, 1, 9, and~10 ordered according to increasing positive contrast. (2) Broad and deep profiles of Clusters~2 and~5, where central maxima and neighboring minima exhibit similar amplitudes in contrast profiles. (3) Contrast profiles, where the central maximum is less pronounced and the contrast is almost everywhere negative, i.e., in Clusters~6, 7, and~8. Only the profiles of Cluster~4 cannot be clearly classified. Their location in the t-SNE projection (right panel of Fig.~\ref{FIG09}) favors Class~2, whereas the predominantly negative contrast in Fig.~\ref{FIG10} suggests Class~3. 

In Fig.~\ref{FIG11}, we present the re-projection of the ten selected clusters in terms of the four CM parameters, i.e., cloud velocity $v$, optical depth $\tau$, source function $S$, and Doppler width $\Delta\lambda_D$. The most conspicuous difference to Fig.~\ref{FIG08} is evident in the Doppler width $\Delta\lambda_D$, which no longer shows a continuous gradient. Instead of that, a vertical line with small Doppler width is now located in the center of the t-SNE map with the largest values to the right and moderately large values to the left. Overall, the projection can be separated in two large clusters with high (Class~1) and low (Classes~2 and~3) values of the source function. The distinction between Classes~2 and~3 is mainly due to differences in the cloud velocities, i.e., blue and redshifts of the spectral line. Closer inspection of the t-SNE projections of CM parameters hint at a sub-class of profiles belonging to Clusters~2 and~5, which reside in a wedge-like structure in the bottom-right corner. In summary, the initial ten clusters can be reduced to two or three classes of spectral profiles, which are associated with chromospheric H$\alpha$ absorption features.
\newpage


\section{Discussion} \label{sec:discussion}

The visual classification of (solar) spectra has a long history and led to a variety of terms describing spectra based on their morphology, e.g., line width, line depression, rest intensity, line gaps, line-wing or line-core emission, satellite components, multi-lobed profiles, etc. Thus, recovering some of these features or combinations thereof with machine learning is expected. However, interpreting the two-dimensional t-SNE maps is still subjective and requires an experienced scientist who brings together the morphology of the spectra and their relation to observed features on the Sun and their underlying physics.

\begin{figure*}
\centering
\includegraphics[width=\textwidth]{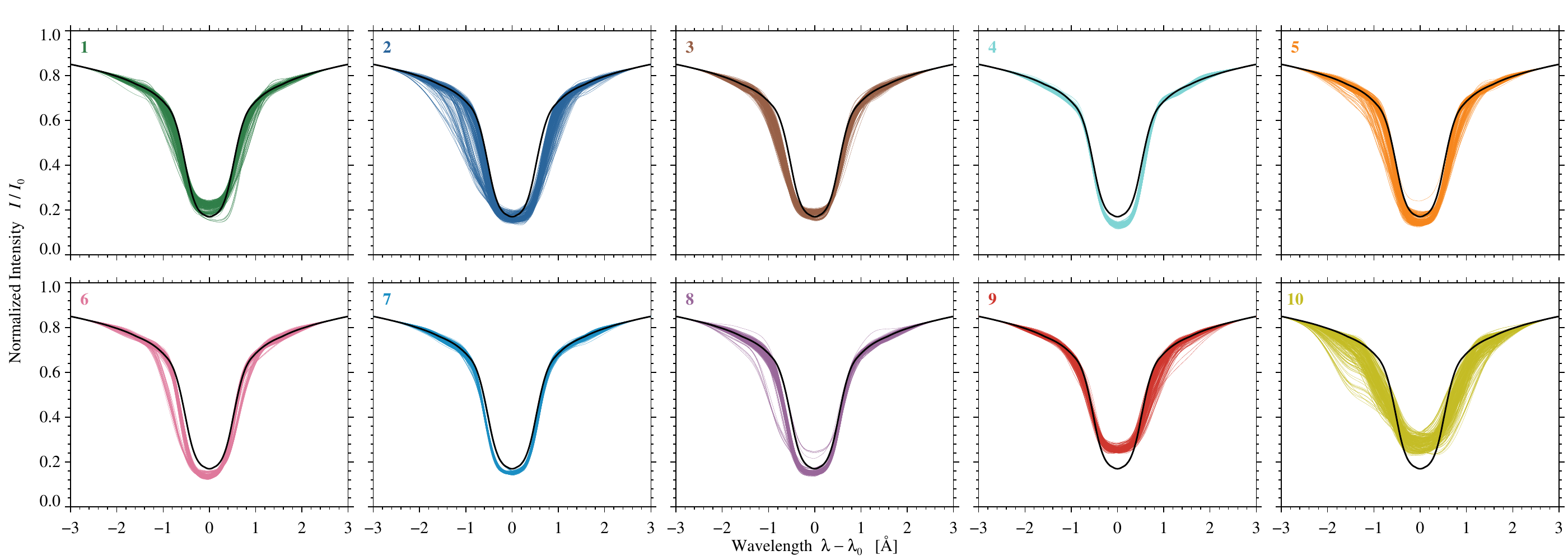}
\includegraphics[width=\textwidth]{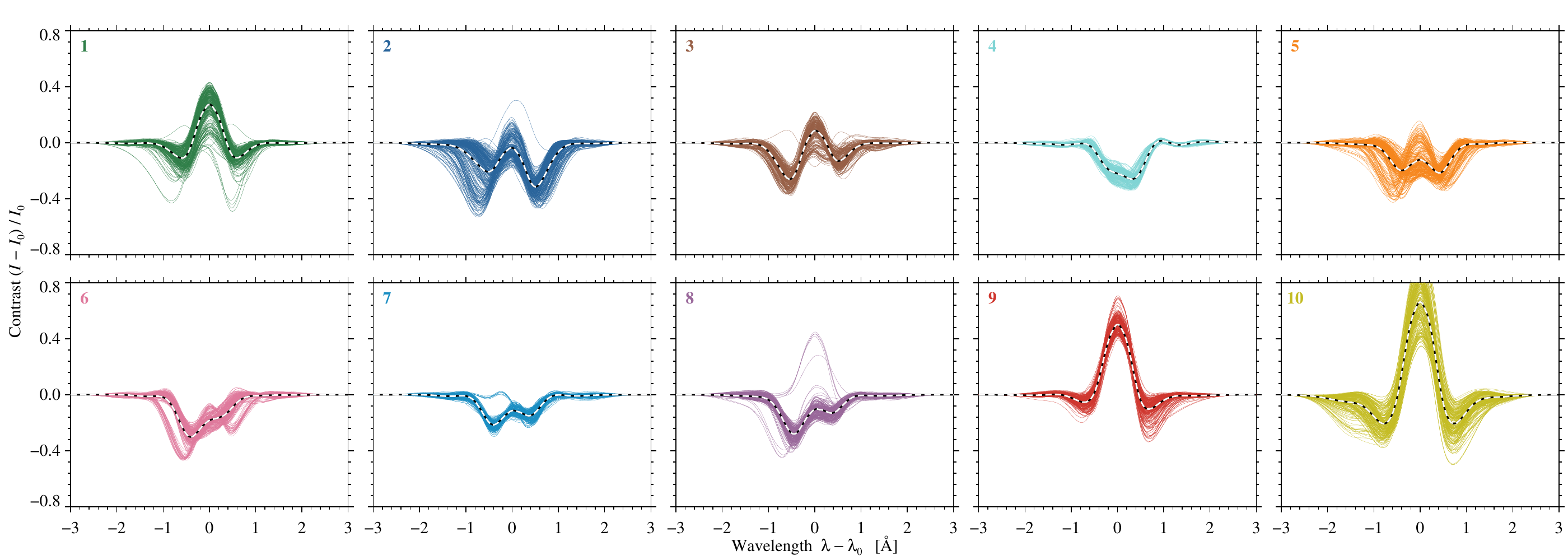}
\caption{Three hundred randomly selected intensity (\textit{top two rows of panels}) and 
    contrast (\textit{bottom two rows of panels}) profiles for the ten clusters of spectra. The color code is the same as in Fig.~\ref{FIG09}. The average contrast profile (\textit{black-white dashed}) for each cluster is plotted to capture the essential shape of the profiles. The quiet-Sun intensity profile (\textit{black}) is plotted for reference.}
\label{FIG10}
\end{figure*}

t-SNE is not the only machine learning algorithm, which can be used for clustering or classification of spectral data. We inspected another method, i.e., Uniform Manifold Approximation and Projection \citep[UMAP,][]{McInnes2018},\footnote{\href{https://github.com/lmcinnes/umap/}{github.com/lmcinnes/umap}} which belongs to the class of k-neighbour-based graph learning algorithms. It provides non-linear dimensionality reduction and furnishes visualization of patterns or clusters in the data without having \textit{a priori} knowledge of the labels in the data. The foundation of UMAP is largely based on manifold theory and topological data analysis. Initial results of UMAP (not presented here) indicate a very good performance in identifying profiles suitable for CM inversions. A full evaluation of UMAP will add another layer of complexity and is thus beyond the scope of this work. However, we see its potential when dealing with larger datasets encompassing observations of different features of the active chromosphere.

Only 27.9\% ($n_\mathrm{CM} =  116\,363$) of the profiles $n_s$ in the best scan produce reliable CM inversions, i.e., they have high linear and rank-order correlation coefficients between observed and CM-inverted contrast profiles (see Fig.~\ref{FIG08}). The t-SNE projection with hexagonal bins is in principle a two-dimensional histogram. Each hexagonal bin contains a certain number of profiles, which are then averaged as demonstrated in Fig.~\ref{FIG02}. Applying a threshold to the suitability parameter (Sect.~\ref{SEC44}) yields significantly less profiles for the ten clusters in Fig.~\ref{FIG09}, i.e., only 8.9\% of $n_s$ ($n_{10} = 37\,033$). Computing the intersection of $n_\mathrm{CM} \cap n_{10} = 36\,392$ indicates that less than 1000 profiles, which were previously deemed unsuitable for CM inversions, share properties common to all profiles in a specific cluster. Selection of these profiles should not be considered as a misclassification but result from the subjective thresholds applied to the correlation coefficients and to the representation of the suitability parameter as a real number (e.g., Fig.~\ref{FIG02}).


\begin{figure*}
\centering
\includegraphics[width=0.24\textwidth]{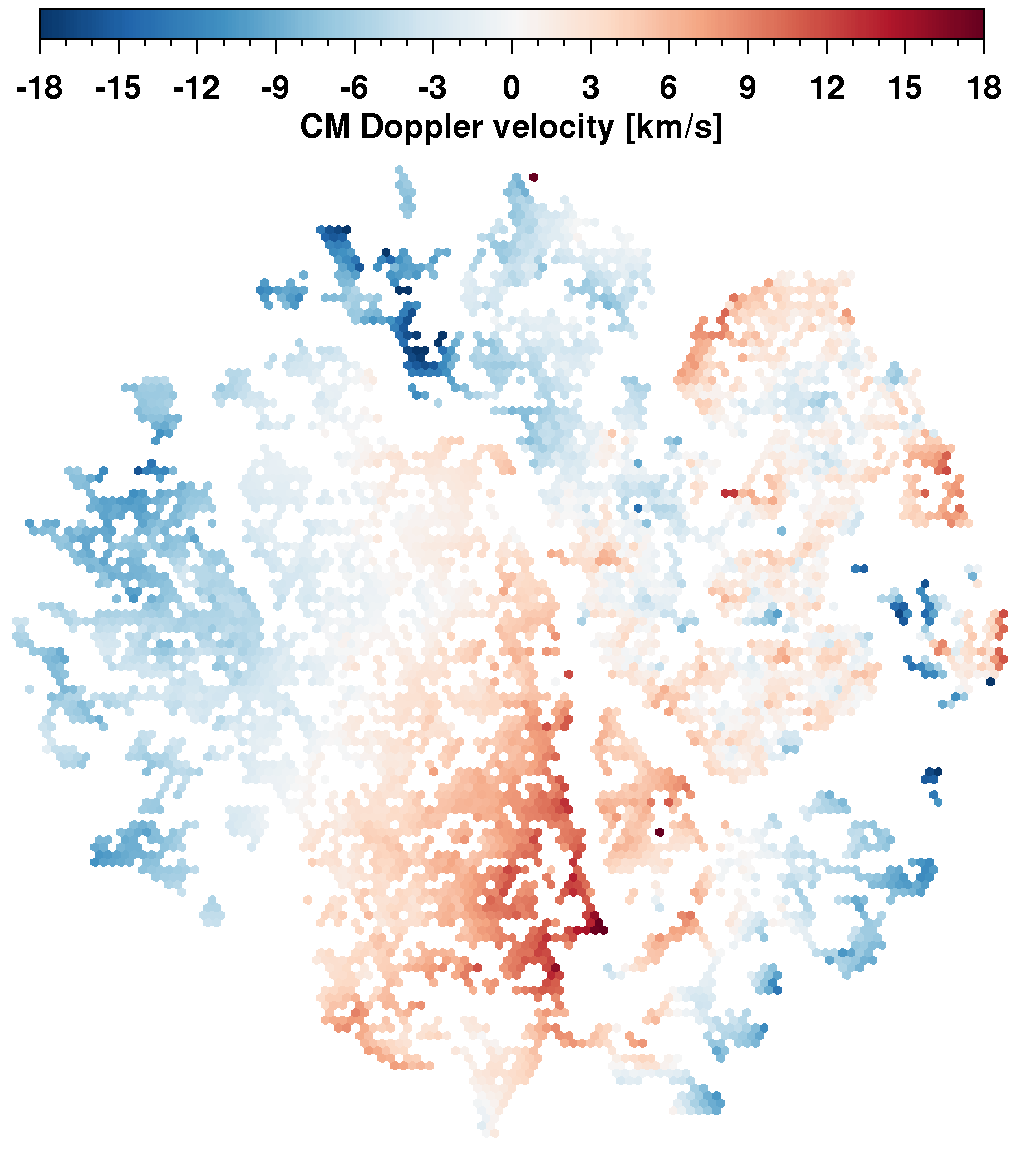}
\includegraphics[width=0.24\textwidth]{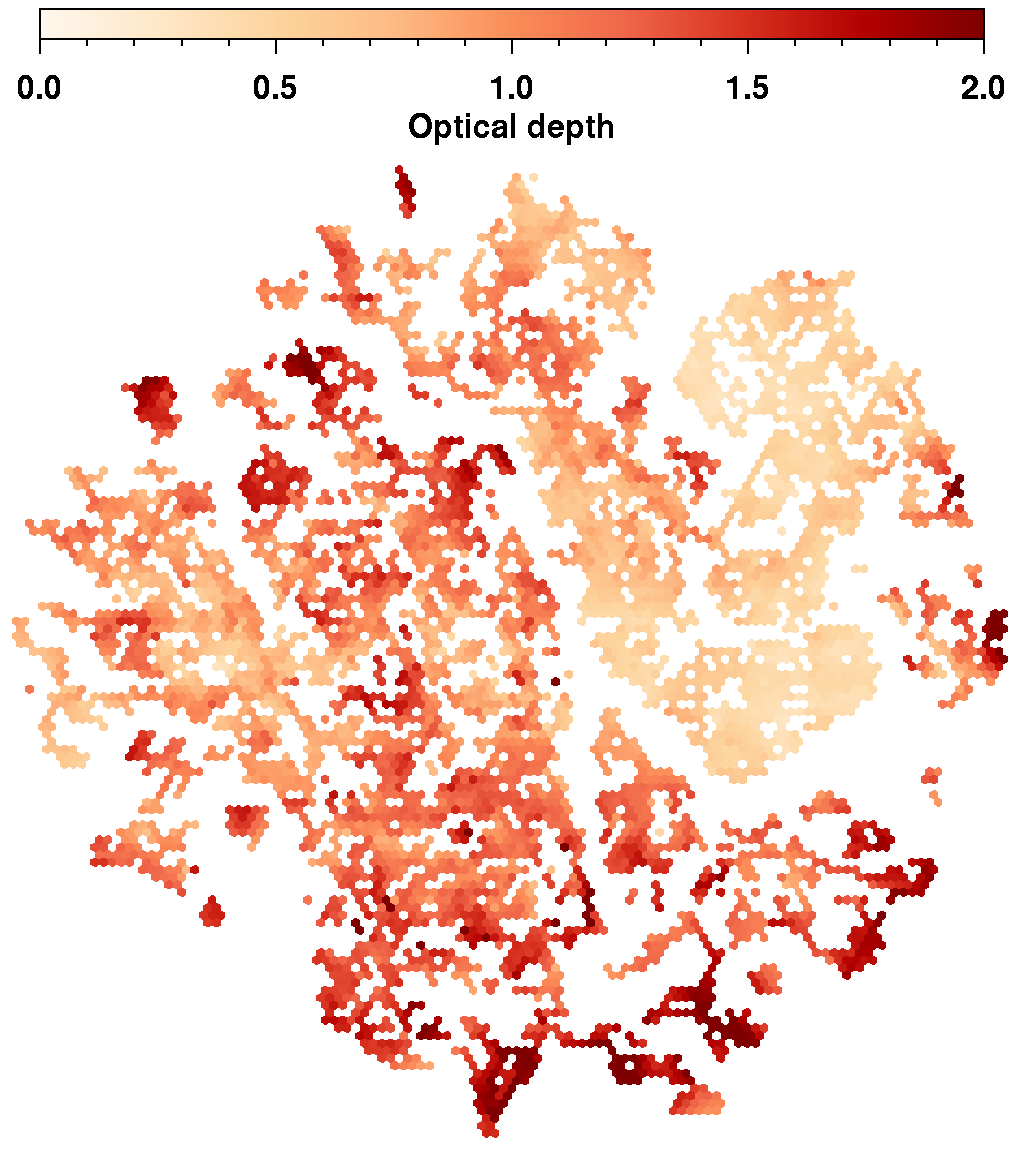}
\includegraphics[width=0.24\textwidth]{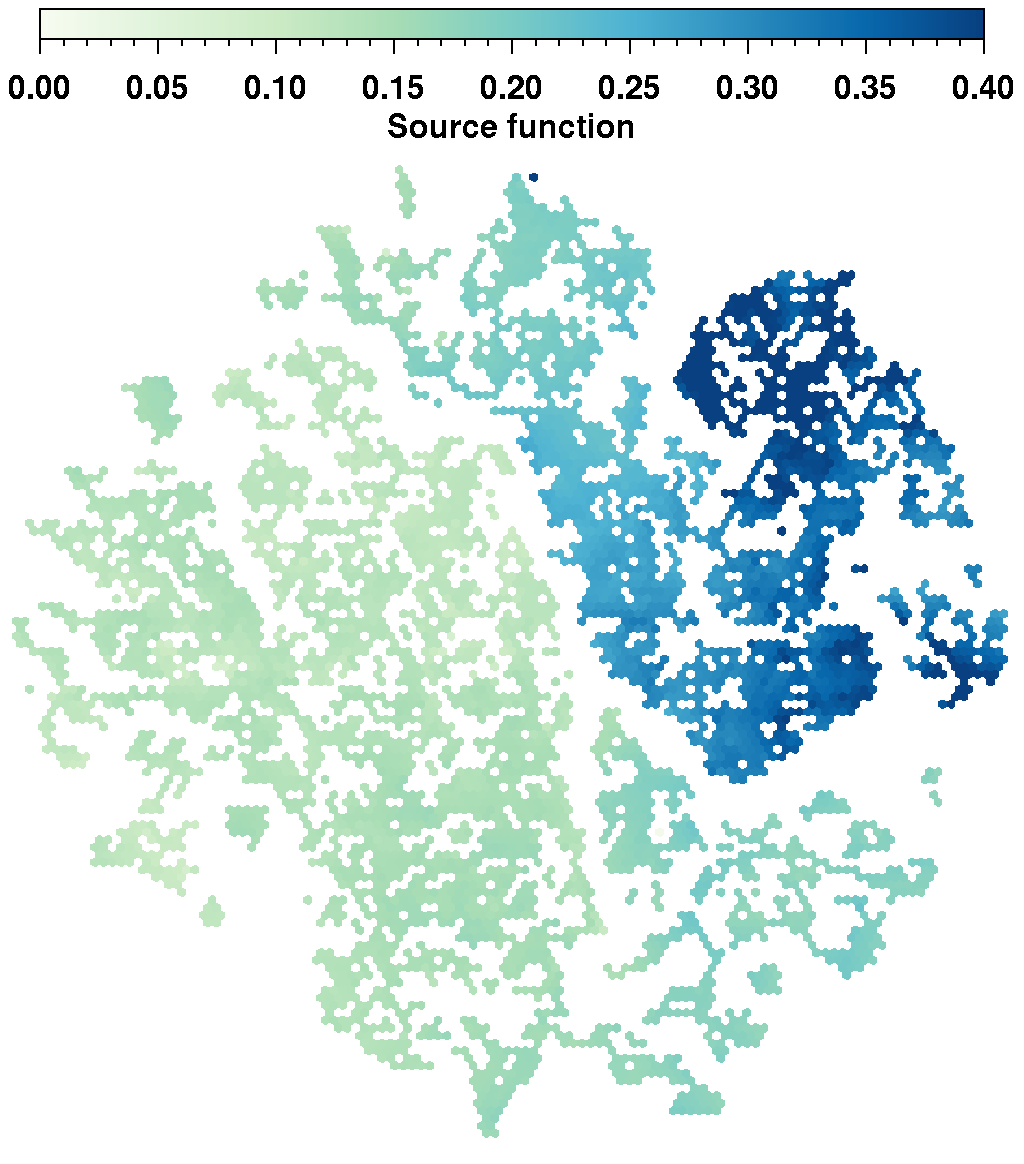}
\includegraphics[width=0.24\textwidth]{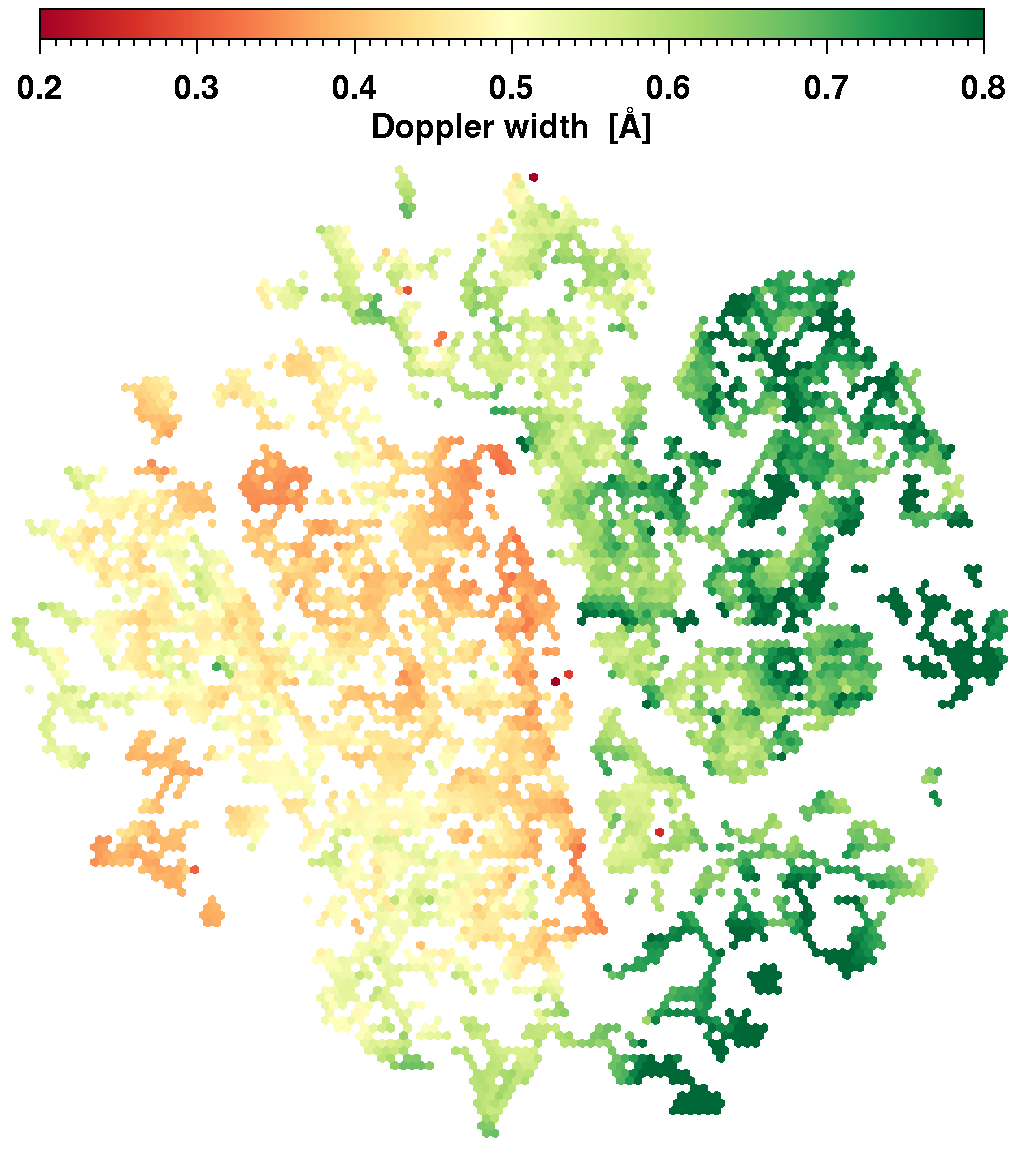}
\caption{Contrast profiles identified in Fig.~\ref{FIG09} yield these t-SNE
    projections of the four CM parameters: cloud velocity $v$, optical depth $\tau$, source function $S$, and Doppler width $\Delta\lambda_D$ (\textit{from left to right}). The range of the parameters is same as in Fig.~\ref{FIG08}.}
\label{FIG11}
\end{figure*}

Even though it is not possible to use t-SNE directly for CM inversions, because t-SNE performs a non-parametric mapping of input data, reliably identifying quiet-Sun profiles already significantly reduces the computing time of the CM inversions depending on the quiet-Sun fraction within the FOV. However, \citet{vanderMaaten2009} suggests `parametric' t-SNE as an enhancement, which implements a regressor that minimizes the t-SNE loss function directly so that new data can be incorporated. Thus, new spectral profiles can be inverted using representative CM parameters from its neighborhood in the two-dimensional t-SNE projections as a starting point. This will speed up the iterative part of CM inversions \citep[see][]{Dineva2020} and prevents the Levenberg–Marquardt least-squares minimization scheme \citep{Markwardt2009} getting trapped in local minima. Currently, the CM inversions as implemented in \citet{Dineva2020} take about five hours on an AMD Opteron 6378 processor for a single spatio-spectral data cube. Only some IDL built-in functions and procedures use multi-threading in the CM inversion, i.e., other means of parallel computing are not employed -- in contrast to the multi-core implementation of t-SNE. Thus, computing times for t-SNE projections and CM inversions are not directly comparable and are only given for reference. Parametric t-SNE and other approaches of embedding new data into existing t-SNE projections, which promise further computational gains, will be explored in forthcoming studies with even more diverse spectral profiles and larger datasets.

In \citet{Verma2020b}, we used the same data to study the homologous surge activity related to the continuous flux emergence and strong proper motions in the region. We discussed the surge in terms of changes in high-resolution noise-stripped spectral and contrast profiles and followed the changes in higher atmospheric layers using UV and EUV data from  the Solar Dynamics Observatory \citep[SDO,][]{Pesnell2012}. We noticed that the surge is not a single entity but consists of regions with different optical thickness and velocity values \citep[see Fig.~11 in][]{Verma2020b}. Similar structures are revealed in the two-dimensional line-core intensity map with ten cluster back-projected (middle panel of Fig.~\ref{FIG09}). The surge is located at $x = [84\arcsec,\, 100\arcsec]$ and $y = [34\arcsec,\, 64\arcsec]$. The different parts of the surge are traced by different clusters. The tip of the surge is traced by Cluster~4, the middle part is marked by Cluster~5 with outside edges covered by Cluster~8, whereas the lower part is traced by Cluster~2 interleaved with Cluster~5. The distinctive shape of contrast profiles for each of the mentioned cluster is indicative of different plasma properties. \citet{NobregaSiverio2016} discerned four different plasma populations in a surge in their 2.5-dimensional experiment simulating the emergence of magnetized plasma through solar atmosphere. To perform a one-to-one comparison of the four Clusters 2, 4, 5, and 8 indicated in the surge with numerical modelling is beyond the scope of this work. However, the t-SNE projection unveiled the composite nature of surge plasma.

The goal of this study was to establish a framework for classifying 
high-resolution spectra using t-SNE. Such a framework is particularly relevant for Big Data, not only because of the ease of its application but also because of huge databases attached to solar research infrastructures. The next step is to apply the presented framework to space data \citep[e.g., the Hinode Spectro-Polarimeter (SP),][]{Tsuneta2008}. First, space data are not affected by seeing. Second, SP data comprises high-spatial resolution full-Stokes spectra, adding polarization properties as another dimension. Science cases for t-SNE applications include, for example, identifying profiles belonging to shear flows or selecting penumbral filaments carrying the Evershed flow. In addition, synoptic observations with moderate spatial and spectral resolution, e.g., vectormagnetogram data of the SDO offer another treasure trove yet to be explored by machine learning. Although the full-Stokes spectra cover only a few wavelength points in the spectral line, it should be possible to identify and classify unique profiles belonging to solar features. The upcoming Chinese spacecraft Solar H$\alpha$ Imaging Spectrometer \citep[SHIS,][]{Chen2018}, which recently received approval will furnish H$\alpha$ spectra without seeing contamination. Thus it is ideally suited for testing and expanding our t-SNE based framework. Commissionig the 4-meter Daniel K.\ Inouye Solar Telescope \citep[DKIST,][]{Tritschler2016} is imminent and its comprehensive instrument suite will produce a plethora of data including high-spectral, -spatial, and -temporal resolution spectropolarimeteric data covering multiple spectral lines.

Although t-SNE proves to be efficient in clustering high-dimensional data, human inference is required at each step to interpret the results. This study provides a framework and ideas how to tailor a classification scheme towards specific spectral data and science questions. The level of complexity increases when moving from the simple question if spectral profiles can be successfully inverted to problems of relating classes of spectra to certain features of chromospheric activity. This exploratory work establishes t-SNE as a suitable tool to cluster high-resolution H$\alpha$ spectra but also demonstrates that unsupervised machine learning algorithms provide the means to explore the ever increasing data volume in solar spectroscopy.


\section{Conclusions} \label{sec:conclusions}

In the present work, we exploited the capabilities of t-SNE for classifying high-resolution H$\alpha$ spectra. We answered the following questions: How does t-SNE depend on Barnes-Hut parameter $\theta$, perplexity $p$, and number of iterations $n$? Do t-SNE projections differ for good and bad seeing data? What is the best choice for input data to be passed to the t-SNE algorithm (i.e., preprocessing and initial dimensionality reduction)? Is t-SNE ``as is'' sufficient to recover meaningful physical properties or are iterations and combinations with other techniques needed?

As an unsupervised machine learning algorithm t-SNE is capable, without any 
\textit{a priori} knowledge, to identify (clusters of) H$\alpha$ intensity or contrast profiles, which are suitable for CM inversions. Depending on the preponderance of quiet-Sun profiles, profiles characteristic of the active chromosphere were mainly pushed to the periphery of the t-SNE projection. The detailed parameter study revealed that the default values of Barnes-Hut parameter $\theta=0.5$, perplexity $p = 50$, and number of iterations $n = 1000$  yield already very good results while being computationally efficient. Furthermore, the projections were in general comparable irrespective of the input data used. However, both noise-stripped contrast profiles and their PCA decomposition based on ten eigenvectors performed best. We devised different approaches for classification of H$\alpha$ spectra using either the full dataset or selected profiles, which were suitable for CM inversions. Extracting clusters can be based on this latter criterion or on spectral line properties. Spectral classes can be defined based on t-SNE projections of CM parameter, and they can be validated by back-projection on slit-reconstructed images or maps of other physical parameters. Human inference is an essential part of the classification, which allowed us to reduce the initial number of 10 clusters to two or three classes of H$\alpha$ intensity and contrast profiles. The exact number of classes will depend on the observed scene on the Sun and on the science questions and goals.

In this exploratory study, t-SNE proofed to be a valuable tool for spectral classification and spectral inversions. The large volume of spectroscopic data provided by major solar research infrastructures such as Hinode/SP and IRIS and in the near future DKIST and SHIS confronts solar physics with a data challenge. Machine learning algorithms such as t-SNE are essential in overcoming limitations in analyzing bulk data. Developing own machine learning codes for solar physics applications may not be the right path, considering that much of the progress in machine learning and artificial intelligence arises in the commercial sector. However, awareness of this rapidly developing sector and adaptation of existing codes promises expedient scientific returns.
\vspace{5mm}


\acknowledgments The Vacuum Tower Telescope (VTT) at the Spanish Observatorio del Teide of the Instituto de Astrof\'{\i}sica de Canarias is operated by the German consortium of the Leibniz-Institut f\"ur Sonnenphysik in Freiburg, the Leibniz-Institut f\"ur Astrophysik Potsdam, and the Max-Planck-Institut f\"ur Sonnensystemforschung in G\"ottingen. This study was supported by grants DE~787/5-1 (CD, CK, IK, MV) and VE~1112/1-1 (MV) of the Deutsche Forschungsgemeinschaft (DFG) and by European Commissions Horizon 2020 Program under grant agreements 824064 (ESCAPE -- European Science Cluster of Astronomy \& Particle Physics ESFRI Research Infrastructures) and 824135 (SOLARNET -- Integrating High Resolution Solar Physics).
\vspace{5mm}

\facilities{Vacuum Tower Telescope (echelle spectrograph)}
\vspace{5mm}

\software{%
    Interactive Data Language (IDL),
    MPFIT \citep{Markwardt2009},
    Python,
    SolarSoft \citep{Bentley1998, Freeland1998}, 
    sTools \citep{Kuckein2017},
    t-SNE \citep{vanderMaaten2008}}



\begin{thebibliography}{}
\expandafter\ifx\csname natexlab\endcsname\relax\def\natexlab#1{#1}\fi
\providecommand{\url}[1]{\href{#1}{#1}}
\providecommand{\dodoi}[1]{doi:~\href{http://doi.org/#1}{\nolinkurl{#1}}}
\providecommand{\doeprint}[1]{\href{http://ascl.net/#1}{\nolinkurl{http://ascl.net/#1}}}
\providecommand{\doarXiv}[1]{\href{https://arxiv.org/abs/#1}{\nolinkurl{https://arxiv.org/abs/#1}}}

\bibitem[{{Anders} {et~al.}(2018){Anders}, {Chiappini}, {Santiago},
  {Matijevi{\v c}}, {Queiroz}, {Steinmetz}, \& {Guiglion}}]{Anders2018}
{Anders}, F., {Chiappini}, C., {Santiago}, B.~X., {et~al.} 2018, \aap, 619,
  A125, \dodoi{10.1051/0004-6361/201833099}

\bibitem[{{Asensio Ramos} \& {D{\'\i}az Baso}(2019)}]{AsensioRamos2019}
{Asensio Ramos}, A., \& {D{\'\i}az Baso}, C.~J. 2019, \aap, 626, A102,
  \dodoi{10.1051/0004-6361/201935628}

\bibitem[{{Barnes} \& {Hut}(1986)}]{Barnes1986}
{Barnes}, J., \& {Hut}, P. 1986, \nat, 324, 446

\bibitem[{{Beckers}(1964)}]{Beckers1964}
{Beckers}, J.~M. 1964, PhD thesis, University of Utrecht

\bibitem[{{Bentley} \& {Freeland}(1998)}]{Bentley1998}
{Bentley}, R.~D., \& {Freeland}, S.~L. 1998, in ESA Special Publication, Vol.
  417, Crossroads for European Solar and Heliospheric Physics. Recent
  Achievements and Future Mission Possibilities, 225--228

\bibitem[{{Carroll} \& {Kopf}(2008)}]{Carroll2008}
{Carroll}, T.~A., \& {Kopf}, M. 2008, \aap, 481, L37,
  \dodoi{10.1051/0004-6361:20079197}

\bibitem[{{Carroll} \& {Staude}(2001)}]{Carroll2001}
{Carroll}, T.~A., \& {Staude}, J. 2001, \aap, 378, 316,
  \dodoi{10.1051/0004-6361:20011167}

\bibitem[{{Chen}(2018)}]{Chen2018}
{Chen}, P.~F. 2018, {Sci. China Phys. Mech. Astron.}, 61, 109631,
  \dodoi{10.1007/s11433-018-9282-y}

\bibitem[{{David}(1961)}]{David1961}
{David}, K.-H. 1961, \zap, 53, 37

\bibitem[{{De Pontieu} {et~al.}(2014){De Pontieu}, {Title}, {Lemen}, {Kushner},
  {Akin}, {Allard}, {Berger}, {Boerner}, {Cheung}, {Chou}, {Drake}, {Duncan},
  {Freeland}, {Heyman}, {Hoffman}, {Hurlburt}, {Lindgren}, {Mathur}, {Rehse},
  {Sabolish}, {Seguin}, {Schrijver}, {Tarbell}, {W{\"u}lser}, {Wolfson},
  {Yanari}, {Mudge}, {Nguyen-Phuc}, {Timmons}, {van Bezooijen}, {Weingrod},
  {Brookner}, {Butcher}, {Dougherty}, {Eder}, {Knagenhjelm}, {Larsen},
  {Mansir}, {Boyle}, {Cheimets}, {DeLuca}, {Gates}, {Hertz}, {McKillop},
  {Park}, {Perry}, {Podgorski}, {Reeves}, {Testa}, {Tian}, {Weber}, {Dunn},
  {Eccles}, {Jaeggli}, {Kankelborg}, {Mashburn}, {Pust}, {Springer},
  {Carvalho}, {Kleint}, {Marmie}, {Mazmanian}, {Pereira}, {Sawyer}, {Strong},
  {Worden}, {Carlsson}, {Leenaarts}, {Wiesmann}, {Chu}, {Bush}, {Scherrer},
  {Martinez-Sykora}, {Lites}, {McIntosh}, {Uitenbroek}, {Okamoto}, {Gummin},
  {Auker}, {Jerram}, {Pool}, \& {Waltham}}]{DePontieu2014}
{De Pontieu}, B., {Title}, A.~M., {Lemen}, J.~R., {et~al.} 2014, \solphys, 289,
  2733

\bibitem[{{De Silva} {et~al.}(2015){De Silva}, {Freeman}, {Bland-Hawthorn},
  {Martell}, {de Boer}, {Asplund}, {Keller}, {Sharma}, {Zucker}, {Zwitter},
  {Anguiano}, {Bacigalupo}, {Bayliss}, {Beavis}, {Bergemann}, {Campbell},
  {Cannon}, {Carollo}, {Casagrande}, {Casey}, {Da Costa}, {D'Orazi}, {Dotter},
  {Duong}, {Heger}, {Ireland}, {Kafle}, {Kos}, {Lattanzio}, {Lewis}, {Lin},
  {Lind}, {Munari}, {Nataf}, {O'Toole}, {Parker}, {Reid}, {Schlesinger},
  {Sheinis}, {Simpson}, {Stello}, {Ting}, {Traven}, {Watson}, {Wittenmyer},
  {Yong}, \& {{\v Z}erjal}}]{DeSilva2015}
{De Silva}, G.~M., {Freeman}, K.~C., {Bland-Hawthorn}, J., {et~al.} 2015,
  \mnras, 449, 2604, \dodoi{10.1093/mnras/stv327}

\bibitem[{{Delgado Mena} {et~al.}(2017){Delgado Mena}, {Tsantaki}, {Adibekyan},
  {Sousa}, {Santos}, {Gonz{\'a}lez Hern{\'a}ndez}, \&
  {Israelian}}]{DelgadoMena2017}
{Delgado Mena}, E., {Tsantaki}, M., {Adibekyan}, V.~Z., {et~al.} 2017, \aap,
  606, A94, \dodoi{10.1051/0004-6361/201730535}

\bibitem[{{Deng} {et~al.}(2015){Deng}, {Zhang}, {Wang}, {Ji}, {Wang}, {Liu},
  {Xiang}, {Jin}, \& {Cao}}]{Deng2015}
{Deng}, H., {Zhang}, D., {Wang}, T., {et~al.} 2015, \solphys, 290, 1479,
  \dodoi{10.1007/s11207-015-0676-1}

\bibitem[{{Denker} {et~al.}(2018){Denker}, {Dineva}, {Balthasar}, {Verma},
  {Kuckein}, {Diercke}, \& {Gonz{\'a}lez Manrique}}]{Denker2018b}
{Denker}, C., {Dineva}, E., {Balthasar}, H., {et~al.} 2018, \solphys, 293, 44,
  \dodoi{10.1007/s11207-018-1261-1}

\bibitem[{{Dineva} {et~al.}(2020){Dineva}, {Verma}, {Gonz\'alez Manrique},
  {Schwartz}, \& {Denker}}]{Dineva2020}
{Dineva}, E., {Verma}, M., {Gonz\'alez Manrique}, S.~J., {Schwartz}, P., \&
  {Denker}, C. 2020, Astron. Nachr., 341, 64, \dodoi{10.1002/asna.202013652}

\bibitem[{{Ester} {et~al.}(1996){Ester}, {Kriegel}, {Sander}, \&
  {Xu}}]{Ester1996}
{Ester}, M., {Kriegel}, H.-P., {Sander}, J., \& {Xu}, X. 1996, in Second
  Conference on Knowledge Discovery and Data Mining, ed. E.~{Simoudis},
  J.~{Han}, \& U.~{Fayyad}, Assoc. Adv. Artif. Intell., 226--231

\bibitem[{{Freeland} \& {Handy}(1998)}]{Freeland1998}
{Freeland}, S.~L., \& {Handy}, B.~N. 1998, \solphys, 182, 497,
  \dodoi{10.1023/A:1005038224881}

\bibitem[{{Hinton} \& {Roweis}(2002)}]{Hinton2002}
{Hinton}, G.~E., \& {Roweis}, S.~T. 2002, in Adv. Neur. Info. Proc. Sys.,
  Vol.~15, In Advances in Neural Information Processing Systems, ed.
  S.~{Becker}, S.~{Thrun}, \& K.~{Obermayer}, 833--840

\bibitem[{{Kos} {et~al.}(2018){Kos}, {Bland-Hawthorn}, {Freeman}, {Buder},
  {Traven}, {De Silva}, {Sharma}, {Asplund}, {Duong}, {Lin}, {Lind}, {Martell},
  {Simpson}, {Stello}, {Zucker}, {Zwitter}, {Anguiano}, {Da Costa}, {D'Orazi},
  {Horner}, {Kafle}, {Lewis}, {Munari}, {Nataf}, {Ness}, {Reid}, {Schlesinger},
  {Ting}, \& {Wyse}}]{Kos2018}
{Kos}, J., {Bland-Hawthorn}, J., {Freeman}, K., {et~al.} 2018, \mnras, 473,
  4612, \dodoi{10.1093/mnras/stx2637}

\bibitem[{{Kuckein} {et~al.}(2017){Kuckein}, {Denker}, {Verma}, {Balthasar},
  {Gonz{\'a}lez Manrique}, {Louis}, \& {Diercke}}]{Kuckein2017}
{Kuckein}, C., {Denker}, C., {Verma}, M., {et~al.} 2017, in IAU Symp., Vol.
  327, Fine Structure and Dynamics of the Solar Atmosphere, ed. S.~{Vargas
  Dom{\'{\i}}nguez}, A.~G. {Kosovichev}, L.~{Harra}, \& P.~{Antolin}, 20--24

\bibitem[{{Kuckein} {et~al.}(2020){Kuckein}, {Gonz{\'a}lez Manrique}, {Kleint},
  \& {Asensio Ramos}}]{Kuckein2020}
{Kuckein}, C., {Gonz{\'a}lez Manrique}, S.~J., {Kleint}, L., \& {Asensio
  Ramos}, A. 2020, \aap, 640, A71, \dodoi{10.1051/0004-6361/202038408}

\bibitem[{{Kuckein} {et~al.}(2016){Kuckein}, {Verma}, \&
  {Denker}}]{Kuckein2016}
{Kuckein}, C., {Verma}, M., \& {Denker}, C. 2016, \aap, 589, A84

\bibitem[{{Kullback}(1959)}]{Kullback1959}
{Kullback}, S. 1959, {Information Theory and Statistics} (New York: Wiley)

\bibitem[{{Markwardt}(2009)}]{Markwardt2009}
{Markwardt}, C.~B. 2009, in ASP Conf.\ Ser., Vol. 411, Astronomical Data
  Analysis Software and Systems XVIII, ed. D.~A. {Bohlender}, D.~{Durand}, \&
  P.~{Dowler}, 251--254

\bibitem[{{Matijevi{\v c}} {et~al.}(2017){Matijevi{\v c}}, {Chiappini},
  {Grebel}, {Wyse}, {Zwitter}, {Bienaym{\'e}}, {Bland-Hawthorn}, {Freeman},
  {Gibson}, {Gilmore}, {Helmi}, {Kordopatis}, {Kunder}, {Munari}, {Navarro},
  {Parker}, {Reid}, {Seabroke}, {Siviero}, {Steinmetz}, \&
  {Watson}}]{Matijevic2017}
{Matijevi{\v c}}, G., {Chiappini}, C., {Grebel}, E.~K., {et~al.} 2017, \aap,
  603, A19, \dodoi{10.1051/0004-6361/201730417}

\bibitem[{{McInnes} {et~al.}(2018){McInnes}, {Healy}, \&
  {Melville}}]{McInnes2018}
{McInnes}, L., {Healy}, J., \& {Melville}, J. 2018, ArXiv e-prints.
\newblock \doarXiv{1802.03426}

\bibitem[{{N{\'o}brega-Siverio} {et~al.}(2016){N{\'o}brega-Siverio},
  {Moreno-Insertis}, \& {Mart{\'{\i}}nez-Sykora}}]{NobregaSiverio2016}
{N{\'o}brega-Siverio}, D., {Moreno-Insertis}, F., \& {Mart{\'{\i}}nez-Sykora},
  J. 2016, \apj, 822, 18, \dodoi{10.3847/0004-637X/822/1/18}

\bibitem[{{Panos} \& {Kleint}(2020)}]{Panos2020}
{Panos}, B., \& {Kleint}, L. 2020, \apj, 891, 17,
  \dodoi{10.3847/1538-4357/ab700b}

\bibitem[{{Panos} {et~al.}(2018){Panos}, {Kleint}, {Huwyler}, {Krucker},
  {Melchior}, {Ullmann}, \& {Voloshynovskiy}}]{Panos2018}
{Panos}, B., {Kleint}, L., {Huwyler}, C., {et~al.} 2018, \apj, 861, 62,
  \dodoi{10.3847/1538-4357/aac779}

\bibitem[{{Pesnell} {et~al.}(2012){Pesnell}, {Thompson}, \&
  {Chamberlin}}]{Pesnell2012}
{Pesnell}, W.~D., {Thompson}, B.~J., \& {Chamberlin}, P.~C. 2012, \solphys,
  275, 3

\bibitem[{{Sainz Dalda} {et~al.}(2019){Sainz Dalda}, {de la Cruz
  Rodr{\'\i}guez}, {De Pontieu}, \& {Go{\v{s}}i{\'c}}}]{Dalda2019}
{Sainz Dalda}, A., {de la Cruz Rodr{\'\i}guez}, J., {De Pontieu}, B., \&
  {Go{\v{s}}i{\'c}}, M. 2019, \apjl, 875, L18, \dodoi{10.3847/2041-8213/ab15d9}

\bibitem[{{Socas-Navarro}(2005)}]{SocasNavarro2005c}
{Socas-Navarro}, H. 2005, \apj, 621, 545, \dodoi{10.1086/427431}

\bibitem[{{Steinmetz} {et~al.}(2006){Steinmetz}, {Zwitter}, {Siebert},
  {Watson}, {Freeman}, {Munari}, {Campbell}, {Williams}, {Seabroke}, {Wyse},
  {Parker}, {Bienaym{\'e}}, {Roeser}, {Gibson}, {Gilmore}, {Grebel}, {Helmi},
  {Navarro}, {Burton}, {Cass}, {Dawe}, {Fiegert}, {Hartley}, {Russell},
  {Saunders}, {Enke}, {Bailin}, {Binney}, {Bland-Hawthorn}, {Boeche}, {Dehnen},
  {Eisenstein}, {Evans}, {Fiorucci}, {Fulbright}, {Gerhard}, {Jauregi}, {Kelz},
  {Mijovi{\'c}}, {Minchev}, {Parmentier}, {Pe{\~n}arrubia}, {Quillen}, {Read},
  {Ruchti}, {Scholz}, {Siviero}, {Smith}, {Sordo}, {Veltz}, {Vidrih}, {von
  Berlepsch}, {Boyle}, \& {Schilbach}}]{Steinmetz2006}
{Steinmetz}, M., {Zwitter}, T., {Siebert}, A., {et~al.} 2006, \aj, 132, 1645,
  \dodoi{10.1086/506564}

\bibitem[{{Traven} {et~al.}(2017){Traven}, {Matijevi{\v c}}, {Zwitter}, {{\v
  Z}erjal}, {Kos}, {Asplund}, {Bland-Hawthorn}, {Casey}, {De Silva}, {Freeman},
  {Lin}, {Martell}, {Schlesinger}, {Sharma}, {Simpson}, {Zucker}, {Anguiano},
  {Da Costa}, {Duong}, {Horner}, {Hyde}, {Kafle}, {Munari}, {Nataf}, {Navin},
  {Reid}, \& {Ting}}]{Traven2017}
{Traven}, G., {Matijevi{\v c}}, G., {Zwitter}, T., {et~al.} 2017, \apjs, 228,
  24, \dodoi{10.3847/1538-4365/228/2/24}

\bibitem[{{Tritschler} {et~al.}(2016){Tritschler}, {Rimmele}, {Berukoff},
  {Casini}, {Kuhn}, {Lin}, {Rast}, {McMullin}, {Schmidt}, {W{\"o}ger}, \&
  {DKIST Team}}]{Tritschler2016}
{Tritschler}, A., {Rimmele}, T.~R., {Berukoff}, S., {et~al.} 2016, Astron.
  Nachr., 337, 1064, \dodoi{10.1002/asna.201612434}

\bibitem[{{Tsuneta} {et~al.}(2008){Tsuneta}, {Ichimoto}, {Katsukawa}, {Nagata},
  {Otsubo}, {Shimizu}, {Suematsu}, {Nakagiri}, {Noguchi}, {Tarbell}, {Title},
  {Shine}, {Rosenberg}, {Hoffmann}, {Jurcevich}, {Kushner}, {Levay}, {Lites},
  {Elmore}, {Matsushita}, {Kawaguchi}, {Saito}, {Mikami}, {Hill}, \&
  {Owens}}]{Tsuneta2008}
{Tsuneta}, S., {Ichimoto}, K., {Katsukawa}, Y., {et~al.} 2008, \solphys, 249,
  167

\bibitem[{{Tziotziou}(2007)}]{Tziotziou2007}
{Tziotziou}, K. 2007, in ASP Conf.\ Ser., Vol. 368, The Physics of
  Chromospheric Plasmas, ed. P.~{Heinzel}, I.~{Dorotovi{\v{c}}}, \& R.~J.
  {Rutten}, 217--237

\bibitem[{{van der Maaten}(2014)}]{vanderMaaten2014}
{van der Maaten}, L. 2014, J. Mach. Learn. Res., 15, 1

\bibitem[{{van der Maaten} \& {Hinton}(2008)}]{vanderMaaten2008}
{van der Maaten}, L., \& {Hinton}, G. 2008, J. Mach. Learn. Res., 9, 2579

\bibitem[{{van der Maaten}(2009)}]{vanderMaaten2009}
{van der Maaten}, L. J.~P. 2009, in Proc. Mach. Learn. Res., Vol.~5, Artificial
  Intelligence and Statistics, ed. D.~{van Dyk} \& M.~{Welling} (Clearwater
  Beach, Florida: PMLR), 384--391

\bibitem[{{Verma} {et~al.}(2019){Verma}, {Matijevi{\v{c}}}, {Denker},
  {Diercke}, {Kuckein}, {Balthasar}, {Dineva}, {Kontogiannis}, \&
  {Pal}}]{Verma2019}
{Verma}, M., {Matijevi{\v{c}}}, G., {Denker}, C., {et~al.} 2019, in Machine
  Learning in Heliophysics, First ML-Helio conference Amsterdam 2019,
  \href{https://ml-helio.github.io/talks/Verma.pdf}{ml-helio.github.io/talks}

\bibitem[{{Verma} {et~al.}(2020){Verma}, {Denker}, {Diercke}, {Kuckein},
  {Balthasar}, {Dineva}, {Kontogiannis}, {Pal}, \& {Sobotka}}]{Verma2020b}
{Verma}, M., {Denker}, C., {Diercke}, A., {et~al.} 2020, \aap, 639, A19,
  \dodoi{10.1051/0004-6361/201936762}

\bibitem[{{von der L{\"u}he}(1998)}]{vonderLuehe1998}
{von der L{\"u}he}, O. 1998, New Astron.\ Rev., 42, 493,
  \dodoi{10.1016/S1387-6473(98)00060-8}

\bibitem[{Wattenberg {et~al.}(2016)Wattenberg, Vi\'egas, \&
  Johnson}]{Wattenberg2016}
Wattenberg, M., Vi\'egas, F., \& Johnson, I. 2016, Distill,
  \dodoi{10.23915/distill.00002}

\end{thebibliography}
\end{document}